\documentclass[aps,prl,showpacs,amsmath,amssymb,amsfonts,superscriptaddress,lengthcheck,twocolumn,longbibliography]{revtex4-1}

\usepackage{graphicx}
\usepackage{subfigure}
\usepackage{verbatim}
\usepackage{dcolumn}
\usepackage{bm}
\usepackage{epsf}
\usepackage{color}
\usepackage[colorlinks=true,citecolor=blue,linkcolor=blue,urlcolor=blue]{hyperref}
\usepackage{hhline}

\newcommand{\bra}[1]{\left\langle #1\right|}
\newcommand{\ket}[1]{\left|#1\right\rangle}
\newcommand{\braket}[2]{\left\langle #1|#2\right\rangle}

\newcommand{\la}{\left\langle}
\newcommand{\ra}{\right\rangle}
\newcommand{\pd}{\partial}

\newcommand{\e}[1]{\exp{\left(#1\right)}}

\newcommand{\com}[2]{\left[#1,\,#2\right]}

\newcommand{\co}[1]{\cos{\left(#1\right)}}
\newcommand{\si}[1]{\sin{\left(#1\right)}}

\newcommand{\bla}{bla\\bla\\bla\\bla\\bla}

\newcommand{\PRA}{Phys. Rev. A }

\newcommand{\PRL}{Phys. Rev. Lett. }

\newcommand{\mc}[1]{\mathcal{#1}}

\newcommand{\mrm}[1]{\mathrm{#1}}

\begin{document}

\title{Trade-off between speed and cost in shortcuts to adiabaticity}

\author{Steve Campbell}
\affiliation{Centre for Theoretical Atomic, Molecular and Optical Physics, Queen's University Belfast, Belfast BT7 1NN, UK}
\affiliation{Istituto Nazionale di Fisica Nucleare, Sezione di Milano \& Dipartimento di Fisica, Universit{\`a} degli Studi di Milano, Via Celoria 16, 20133 Milan, Italy}

\author{Sebastian Deffner}
\affiliation{Department of Physics, University of Maryland Baltimore County, Baltimore, MD 21250, USA}
\date{\today}

\begin{abstract}
{Achieving effectively adiabatic dynamics is a ubiquitous goal in almost all areas of quantum physics. Here, we study the speed with which a quantum system can be driven when employing transitionless quantum driving. As a main result, we establish a rigorous link between this speed, the quantum speed limit, and the (energetic) cost of implementing such a shortcut to adiabaticity. Interestingly, this link elucidates a trade-off between speed and cost, namely that instantaneous manipulation is impossible as it requires an infinite cost. These findings are illustrated for two experimentally relevant systems -- the parametric oscillator and the Landau-Zener model -- which reveal that the spectral gap governs the quantum speed limit as well as the cost for realizing the shortcut.}
\end{abstract}
\maketitle

A popular saying states that \emph{there ain't no such thing as a free lunch}. Although quite casually formulated, this phrase expresses nothing less but the gist of the second law of thermodynamics, namely that non-ideal processes are always accompanied by the irreversible expense of a thermodynamic resource. Nevertheless, recent research in quantum control and quantum thermodynamics has seen growing popularity of so-called `shortcuts to adiabaticity', i.e., fast processes with the same outcome as an ideal, infinitely slow process \cite{Torrontegui2013}. Such shortcuts are fast processes with suppressed nonequilibrium excess energy \cite{Campisi2011,Deffner2016PRE}, and apparently provide means to circumvent the second law in isolated systems \cite{Acconcia2015a,Acconcia2015,Acconcia2016}. Thus, a variety of techniques has been developed: using dynamical invariants \cite{Chen2010}, inversion of scaling laws \cite{Campo2012}, the fast-forward technique for Schr\"odinger \cite{Masuda2010,Masuda2011,Torrontegui2012,Torrontegui2012a,Masuda2014a,Kiely2015} and Dirac dynamics \cite{Deffner2016}, transitionless quantum driving \cite{Demirplak2003,Demirplak2005,Berry2009,Deffner2014}, classical dissipationless driving \cite{Jarzysnki2013,Patra2016}, optimal protocols from optimal control theory~\cite{Khaneja2005,Chen2011a,Stefanatos2013,Campbell2014,Wu2015,Saberi2014,PolkovnikovArXiv}, optimal driving from properties of the quantum work statistics  \cite{Xiao2014}, `environment' assisted methods \cite{Masuda2014}, using the properties of Lie algebras \cite{Torrontegui2014}, and approximate methods such as linear response theory \cite{Acconcia2015} and fast quasistatic dynamics \cite{Garaot2015}. 

Among this plethora of techniques transitionless quantum driving (TQD) is unique. In its original formulation \cite{Demirplak2003,Demirplak2005,Berry2009} one considers a time-dependent Hamiltonian $H_0(t)$ and constructs an additional counterdiabatic field, $H_1(t)$, such that the joint Hamiltonian $H(t)=H_0(t)+H_1(t)$ drives the dynamics precisely through the adiabatic manifold of $H_0(t)$. Moreover, $H_1(t)$ vanishes by construction in the beginning, $t=0$, and and the end, $t=\tau$, of the finite time process. Thus, if only considering the energy balance, $\la H(\tau)\ra-\la H(0)\ra=\la H_0(\tau)\ra-\la H_0(0)\ra$, implementing such a shortcut to adiabaticity appears to be thermodynamically for free \cite{delCampo2013}.  Even more dramatically, it seems that such an energetically free shortcut to adiabaticity could be implemented for any arbitrarily fast process of arbitrarily short duration $\tau$.

That this almost naive interpretation of TQD cannot be entirely sound has been formalised recently in Ref.~\cite{Zheng2015} where a family of cost functionals are introduced. They are given by the time averaged norm of the counterdiabatic field, $C_t^n=\nu_{t,n}\,\int_0^\tau dt\, ||H_1(t)||^n$, where $\nu_{t,n}$ is a set-up dependent constant and the index of the norm $n$ depends on the nature of the applied fields (see Ref.~\cite{Zheng2015} for a more detailed discussion). Here, we show that the norm plays the most crucial role in defining the cost of driving. Therefore, we remove any set-up dependence by assuming that $\nu_{t,n} = n = 1$. Although insightful, defining a cost ad hoc is not entirely satisfactory. In particular, it is not immediately clear how $C_t^n$ corresponds to expended resources. Moreover, the definition of $C_t^n$ also does not address the rather unsettling impression that TQD could be performed in arbitrarily short times $\tau$. In this letter we resolve both issues by showing that a cost function that depends on the norm of the counterdiabatic term is intimately related to maximal speed of the evolution.

It has been established in virtually all areas of quantum physics \cite{Bekenstein1981,lloyd00,Deffner2010,Giovannetti2011} that the Heisenberg uncertainty relation for energy and time sets a quantum speed limit (QSL) \cite{mandelstam45,Bhattacharyya1983,Pfeifer1993,margolus98,Giovannetti2004}, i.e., a fundamental upper bound on the speed of quantum evolution.  These bounds have been extensively studied for isolated \cite{Barnes2013a,Poggi2013,Hegerfeldt2013,Andersson2014,Deffner2013,richerme14,jurcevic14} and open \cite{Deffner2013PRL,delcampo13,taddei13,deffner14,Zhang2014,Mukherjee2013,Xu2014,Xu2014a,Cimmarusti2015,Mondal2016,Mondal2016PLA} systems. It has been shown that the maximal speed of quantum evolution is given by time averaged norm of the generator of the dynamics \cite{Deffner2013PRL}, which in the case of unitary dynamics and for orthogonal states reduces to the average energy $E$ \cite{Deffner2013}. Thus, the minimal time during which a quantum system can evolve from intial to final state, i.e,  the QSL time $\tau_\mrm{QSL}$ is determined by $\tau_\mrm{QSL}\simeq\hbar/2E$ \cite{Deffner2013,Deffner2013PRL}. Since this QSL is a consequence of fundamental properties of quantum dynamics, it also has to apply to quantum processes facilitating shortcuts to adiabaticity. In other words, the QSL \emph{prohibits} TQD to be performed in arbitrarily short times.

In this letter we show that the cost of TQD \cite{Zheng2015} and the QSL \cite{Deffner2013PRL} are intimately connected. As our main result we rigorously prove a trade-off between the speed and the thermodynamic cost: \emph{the faster one wants to implement a shortcut, the higher is the thermodynamic cost of realizing the quantum process}. We will further illustrate that this insight is not only of theoretical and conceptual interest, but also of practical relevance. To this end, we will analyze two experimentally important systems, namely the parametric harmonic oscillator and the Landau-Zener model. Parametric harmonic oscillators have been shown to be ideal testbeds for quantum thermodynamic relations \cite{Deffner2008,Deffner2010CP,Deffner2013PRE,Gong2014}, which can be easily implemented for instance in ion traps \cite{Huber2008,Abah2012,Rossnagel2014,Rossnagel2016}. The Landau-Zener model, on the other hand, is closely related to the Ising model \cite{Dziarmaga2005,delCampo2012LZ,Johansson2009} and hence instrumental for current technological advancements in quantum annealing \cite{Das2009} such as the DWave machine \cite{Bian2010,Johansson2009,Johnson2011}.

\paragraph*{Preliminaries.}
Consider a time-dependent Hamiltonian $H_0(t)$ with instantaneous eigenvalues $\{\varepsilon_n(t)\}$ and eigenstates $\{\ket{n_t}\}$. In the limit of infinitely slow variation of $H_0(t)$, i.e., the adiabatic limit, no transitions between eigenstates occur \cite{Messiah1966}. Now consider a non-adiabatic parameterization of $H_0(t)$. In this case we can construct a corresponding Hamiltonian $H(t)=H_0(t)+H_1(t)$ such that the adiabatic solution of $H_0(t)$ is an exact solution of the dynamics generated by $H(t)$. It can be shown that~\cite{Demirplak2003,Demirplak2005,Berry2009,Zheng2015}
\begin{equation}
\label{eq01}
H_1(t)= i \hbar \com{\pd_t \ket{n_t}\!\!\bra{n_t}}{\ket{n_t}\!\!\bra{n_t}}.
\end{equation}
Note that computing the counterdiabatic Hamiltonian $H_1(t)$ requires the instantaneous eigenbasis $\ket{n_t}$. Since finding these time-dependent eigenstates can become arbitrarily complicated, hybrid methods have been developed utilizing tools from optimal control theory~\cite{Campbell2014,Wu2015,Saberi2014,PolkovnikovArXiv}.

In Ref.~\cite{Zheng2015} a family of functionals has been proposed to quantify the cost associated with implementing $H_1(t)$. The simplest member of the family is given by the trace norm, $||\cdot||$ \cite{Zheng2015,Santos2015,Santos2016,Coulamy2016},
\begin{equation}
\label{eq02}
C^1_t \equiv C=\int_0^\tau\, dt\, ||H_1(t)||,
\end{equation}
with $\nu_{t,1}=1$. It is easy to see that for a single 2-level spin, $\partial_t C$ is proportional to the average power input~\cite{Zheng2015}, i.e., $H_1(t)$ reduces to an orthogonal, magnetic field. More generally, $C$ can be interpreted as the additional action arising from the counterdiabatic driving. Hence, the relation to the QSL becomes apparent, since loosely speaking the QSL sets a lower bound on the action $E\,\tau_\mrm{QSL}\simeq \hbar/2$ \cite{Deffner2013}.

The QSL is a fundamental upper bound on the rate with which a quantum state can evolve. For our present purposes we are interested in the evolution of pure states under the time-dependent Schr\"odinger equation $i\hbar\, \pd_t \ket{\psi_t}=H(t)\ket{\psi_t}$. It has been shown that in this case the maximal rate of change of the angle between initial and time-evolved state $\mc{L}_t=\arccos\left|\braket{\psi_0}{\psi_t}\right|$ is given by
\begin{equation}
\label{eq03}
\pd_t\mc{L}_t\leq v_\mrm{QSL}\equiv\frac{\left|\epsilon_t\right|}{\hbar\,\co{\mc{L}_t} \si{\mc{L}_t}}\,,
\end{equation}
where $\epsilon_t \!\!= \!\! \| H(t) \rho \|$ with $\rho\!\!=\!\!\ket{\psi_t}\bra{\psi_t}$~\cite{Deffner2013PRL}. From this maximal quantum speed one easily obtains the QSL time \cite{Deffner2013PRL},
\begin{equation}
\label{eq04}
t\geq \tau_\mrm{QSL}\equiv\frac{\hbar}{2 E_\tau} \,\left[\si{\mc{L}_\tau}\right]^2\,,
\end{equation}
where $E_\tau$ is the time-averaged norm of the energy, $E_\tau=1/\tau\,\int_0^\tau\, dt\, |\epsilon_t|$ and $\tau$ is the length of the driving protocol. Note that Eq.~\eqref{eq04} is an expression of the Heisenberg uncertainty principle of energy and time for time-dependent, driven quantum systems~\cite{Deffner2013}.

\paragraph*{Trade-off between speed and cost.}
It is easy to see, that in the case of TQD the instantaneous cost, i.e., the trace norm of the counterdiabatic Hamiltonian, $H_1(t)$, reduces to
\begin{equation}
\label{eq05}
\pd_t C=|| H_1(t)||=\sqrt{\braket{\pd_t n_t}{\pd_t n_t}}\,,
\end{equation}
where we used that $\braket{\pd_t n_t}{n_t}=0$, which is true for all Hamiltonians with entirely discrete eigenvalue spectrum. By further noting that $\ket{\psi_t}=\ket{n_t}$ and $H(t)=H_0(t)+H_1(t)$ with $H_1(t)$ as given in Eq.~\eqref{eq01}, $\epsilon_t$ simply becomes
\begin{equation}
\label{eq06}
\epsilon_t=\sqrt{\varepsilon_n^2(t)+\braket{\pd_t n_t}{\pd_t n_t}}\,,
\end{equation}
where we employed again $\braket{\pd_t n_t}{n_t}=0$.

Substituting Eqs.~\eqref{eq05} and \eqref{eq06} into Eq.~\eqref{eq03} we obtain the maximal speed with which a quantum state can undergo transitionless quantum driving
\begin{equation}
\label{eq07}
v_\mrm{QSL}=\frac{\sqrt{\varepsilon_n^2(t)+\left(\pd_t C\right)^2}}{\hbar\,\co{\mc{L}_t} \si{\mc{L}_t}}\,,
\end{equation}
and the QSL time becomes
\begin{equation}
\label{eq08}
\tau_\mrm{QSL}=\frac{\hbar\tau\,\left[\si{\mc{L}_\tau}\right]^2}{2\int_0^\tau{ dt\,\sqrt{\varepsilon_n^2(t)+\left(\pd_t C\right)^2}}}\,.
\end{equation}
The latter two equations constitute our main results. First, we have related the cost for TQD introduced in Ref.~\cite{Zheng2015} with one of the most fundamental results in modern quantum physics -- the Heisenberg uncertainty principle for energy and time. Second, Eqs.~\eqref{eq07} and \eqref{eq08} express, in a transparent and immediate way, the trade-off of speed and cost of a shortcut to adiabaticity. In particular, Eq.~\eqref{eq07} shows that the faster a quantum system evolves along its adiabatic manifold, the higher is the cost of implementing the shortcut \footnote{Interestingly a similar result was obtained in Ref.~\cite{Santos2015}.  However, the discussion in Ref.~\cite{Santos2015} is based on a different version of the QSL \cite{Deffner2013}, which is governed by the off-diagonal matrix elements and not by the norm of $H_1(t)$ \eqref{eq03}. Therefore, the here discussed trade-off could not be analyzed previously.}. Equation~\eqref{eq08} states that the shorter the time during which a quantum system is driven from initial to final energy eigenstate, the more thermodynamic resources have to be expended \footnote{It is interesting to note that recently a similar relation was shown for work fluctuations induced by using TQD~\cite{Funo2016}.}. The remainder of this analysis is dedicated to two experimentally relevant case studies, which illustrate this trade-off for practical applications.

\paragraph*{Case study 1: Harmonic oscillator.}
The `unperturbed' Hamiltonian of the parametric harmonic oscillator reads,
\begin{equation}
\label{eq09}
H_0(t)=\frac{p}{2m}+\frac{1}{2}m\omega_t^2\, x^2\,.
\end{equation}
For the sake of simplicity we only consider situations in which the system is initially prepared in its ground state,
\begin{equation}
\label{eq10}
\psi_0(x)=\left(\frac{m\omega_0}{\pi\hbar}\right)^{1/4}\,\e{-\frac{m\omega_0\,x^2}{2\hbar}}\,
\end{equation}
where the corresponding energy eigenvalue is $\varepsilon_0(0)=\hbar\omega_0/2$. A straightforward calculation then reveals that the cost of keeping the oscillator in its instantaneous ground state at all times is
\begin{equation}
\label{eq11}
\pd_t C=\left|\frac{\pd_t \omega_t}{\sqrt{8}\omega_t}\right|,
\end{equation}
and the maximal quantum speed reads
\begin{equation}
\label{eq12}
v_\mrm{QSL}=\frac{\sqrt{\left(\hbar\omega_t/2\right)^2+\left(\pd_t \omega_t/\sqrt{8}\omega_t\right)^2}}{\hbar\,\co{\mc{L}_t} \si{\mc{L}_t}}\,.
\end{equation}
Finally, the instantaneous angle can be written as
\begin{equation}
\label{eq13}
\mc{L}_t=\arccos\left(\sqrt{2\sqrt{\omega_0\omega_t}/(\omega_0+\omega_t)}\right)\,.
\end{equation}
Note that the maximal speed of quantum evolution, $v_\mrm{QSL}$, is fully determined by the parameterization of the angular frequency $\omega_t$. Therefore, $v_\mrm{QSL}$ fully characterizes the dynamics, and we can analyze the quantum process without having to solve the time-dependent Schr\"odinger equation.

\begin{figure}[t]
{\bf (a)}\\
\includegraphics[width=0.85\columnwidth]{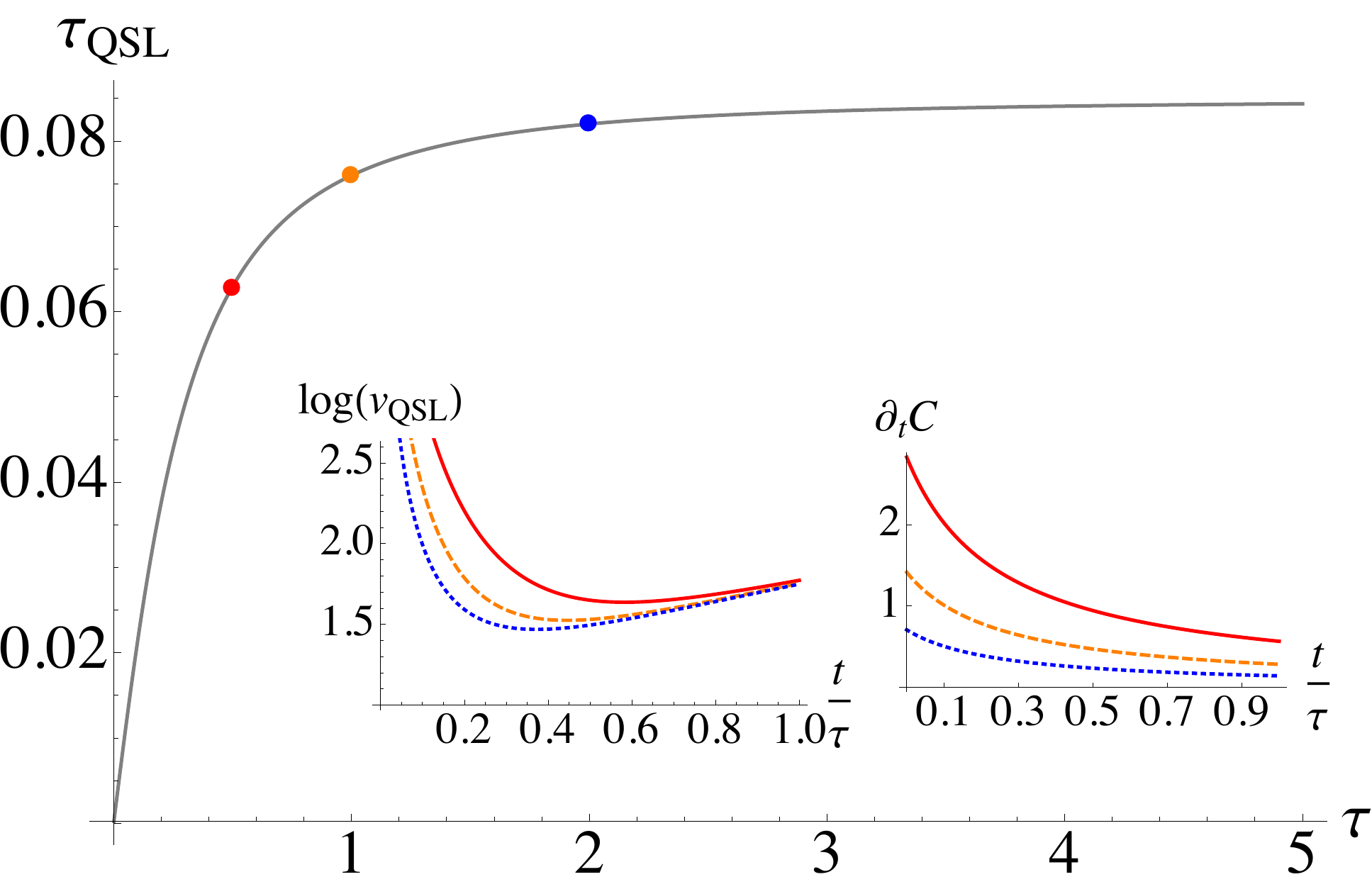}\\
{\bf (b)}\\
\includegraphics[width=0.85\columnwidth]{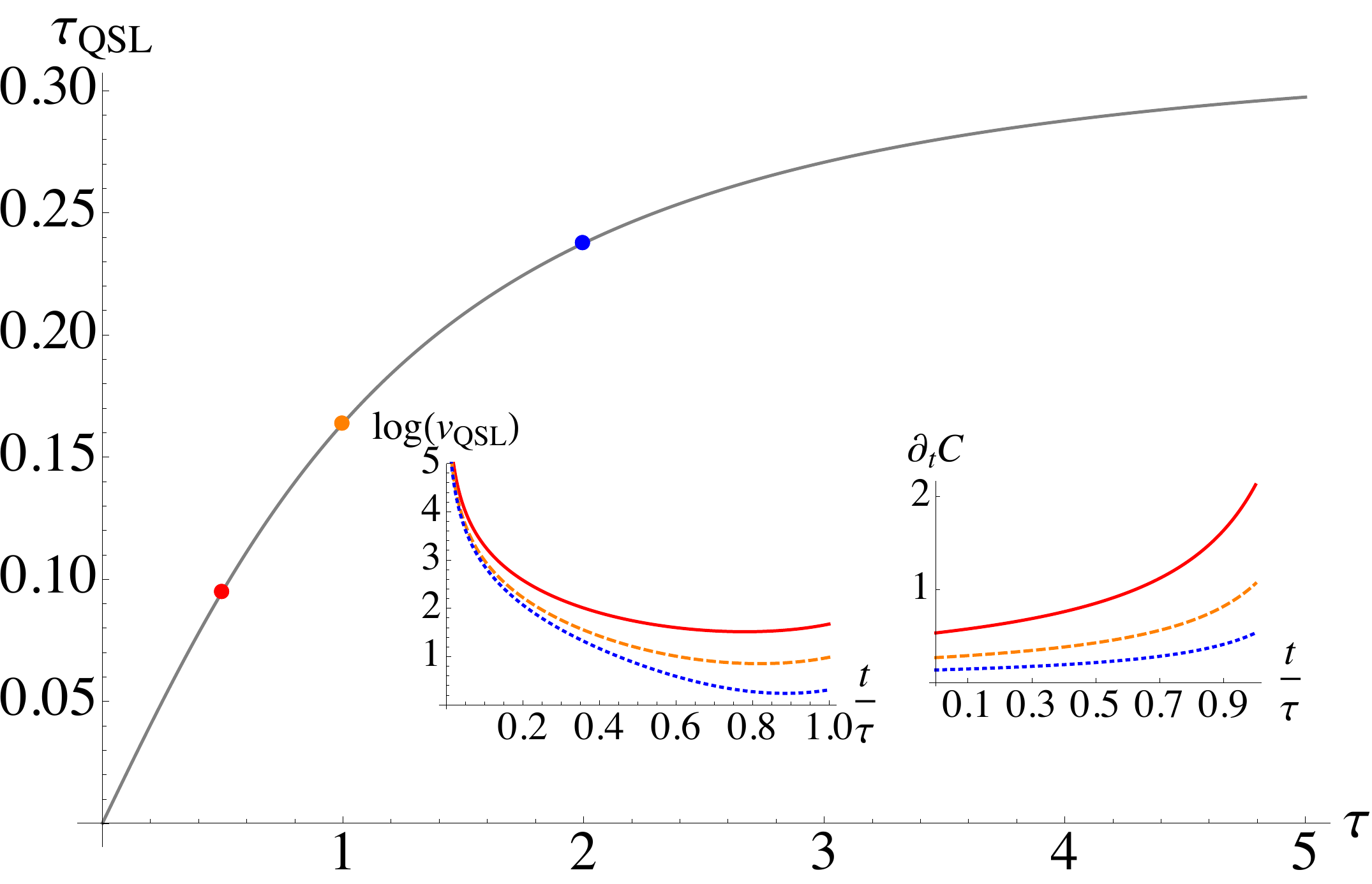}
\caption{We consider the harmonic oscillator with time dependent frequency $\omega_t=\omega_0 + \omega_d \tfrac{t}{\tau}$. {\it Main Panels:} QSL time. The points correspond to the QSL times considered in the insets. {\it Insets:} Maximal speed, $v_\mrm{QSL}$, and instantaneous cost, $\partial_t C$, for $\tau=0.5$ (solid), 1 (dashed) and 2 (dotted). {\bf (a)} A compression with $\omega_0 = 1$ and $\omega_d=4$. {\bf (b)} An expansion with $\omega_0=1$ and $\omega_d=-0.75$.}
\label{fig:HO}
\end{figure}

In Figs.~\ref{fig:HO} we examine the case of a compression [panel {\bf (a)}] and an expansion [panel {\bf (b)}] using the simple linear ramp $\omega_t = \omega_0 + \omega_d (t/\tau)$. Interestingly we see that generally the QSL time is significantly larger for the expansion. Nevertheless, we also observe that using the shortcut can bring the QSL time to arbitrarily small values. However, as evidenced in the insets, smaller $\tau$ corresponds to a diverging instantaneous cost. Remarkably, $v_\mrm{QSL}$ and $\partial_t C$ exhibit qualitatively opposite behaviors for the two protocols. For the compression, we see that $v_\mrm{QSL}$ tends to increase as we decrease the total evolution time $\tau$. However, all curves collapse on top of one another toward the end of the protocol. Conversely, in the case of an expansion the speeds diverge as we evolve. The instantaneous cost qualitatively behaves in the same way. 

This behavior is due to the effect that these protocols have on the energy spectrum. In the case of a compression, the energy levels become more spaced and therefore the first excited state becomes progressively harder to reach. The larger gap then means that we can drive the system comparatively faster without exciting it and the associated cost of achieving this dynamic decreases. In the case of an expansion the energy spacing decreases. Therefore, to avoid excitations, the system must be driven slower. Achieving this dynamics using a shortcut is then necessarily accompanied by an increasingly large cost.

The latter interpretation is further supported by considering Fermi's golden rule for time-dependent perturbation theory~\cite{Messiah1966}. This rule states that the rate of quantum transitions is determined by the time-integrated magnitude of the perturbation. In TQD we seek to suppress these transitions. This means that larger gaps have a lower probability of observing a transition in the `unperturbed' dynamics compared to smaller gaps. Hence, it is also `cheaper' to suppress transitions in processes with larger gaps, than in denser energy spectra.

Clearly, the gap between the driven state and the rest of the spectrum plays the most crucial role in determining both the QSL and the cost of achieving finite time adiabatic dynamics. Such an observation is of particular relevance in critical many-body systems, where quantum phase transitions often occur at avoided crossings in the spectrum. In the following, we examine the avoided crossing (AC) in the Landau-Zener (LZ) model, which serves to elucidate all the relevant features of driving the many-body Ising model through its critical point~\cite{delCampo2012LZ} and is also relevant to the Lipkin-Meshkov-Glick model~\cite{Campbell2014}.

\paragraph*{Case study 2: Landau-Zener model.}
Consider the Hamiltonian
\begin{equation}
\label{eq14}
\mathcal{H}_{LZ} = \Delta \sigma_x + g(t) \sigma_z,
\end{equation}
where $\Delta$ is the energy splitting and $g(t)$ is the time dependent field. As shown in Ref.~\cite{delCampo2012LZ} the Ising model can be expressed as a series of independent LZ crossings, and therefore the following results extrapolate to  driving a critical many-body system. For the sake of clarity we further rescale $\mathcal{H}_{LZ}$ by $\Delta$, $H_0=\mathcal{H}_{LZ}/\Delta$, and hence set the minimal energy gap to 1. The corresponding correction term Eq.~\eqref{eq01} is readily determined to be~\cite{Berry2009}
\begin{equation}
\label{eq15}
H_1 = -\frac{g'(t) \Delta}{2\left(\Delta^2 + [g(t)]^2 \right)} \sigma_y,
\end{equation}
which allows us to evaluate Eqs.~\eqref{eq05} and \eqref{eq07}. 

\begin{figure}[t]
{\bf (a)}\\
\includegraphics[width=0.85\columnwidth]{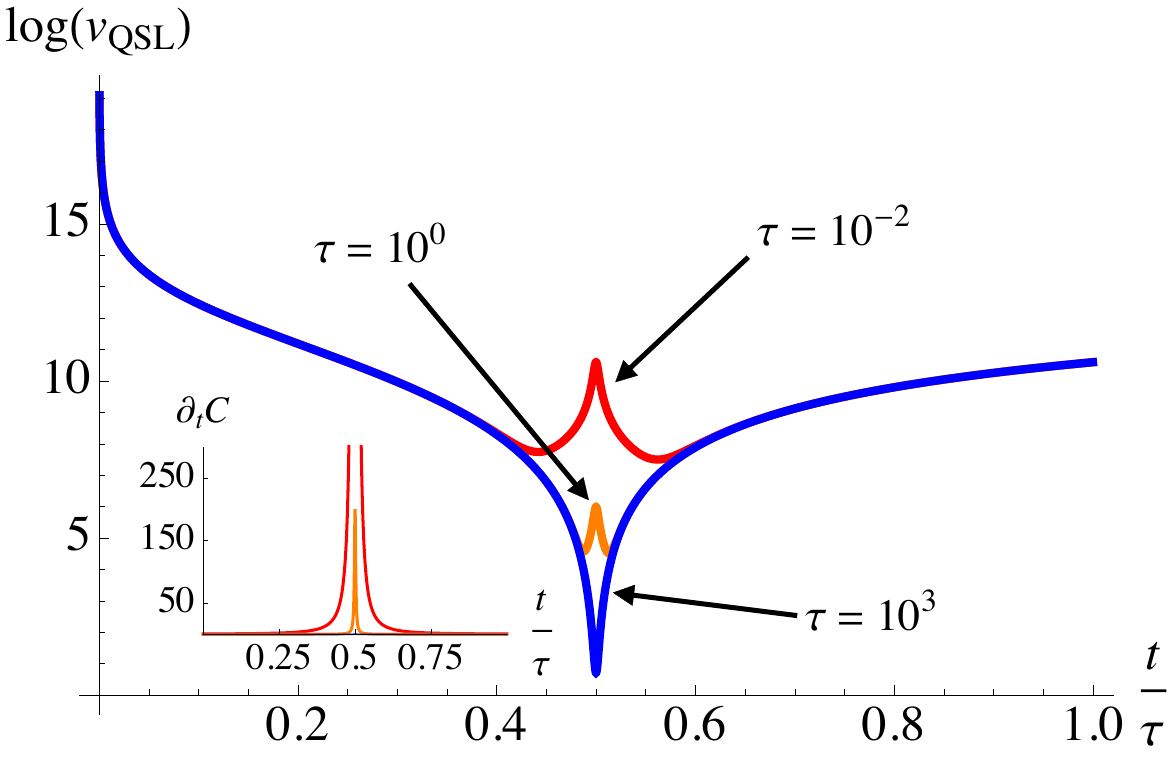}\\
{\bf (b)}\\
\includegraphics[width=0.85\columnwidth]{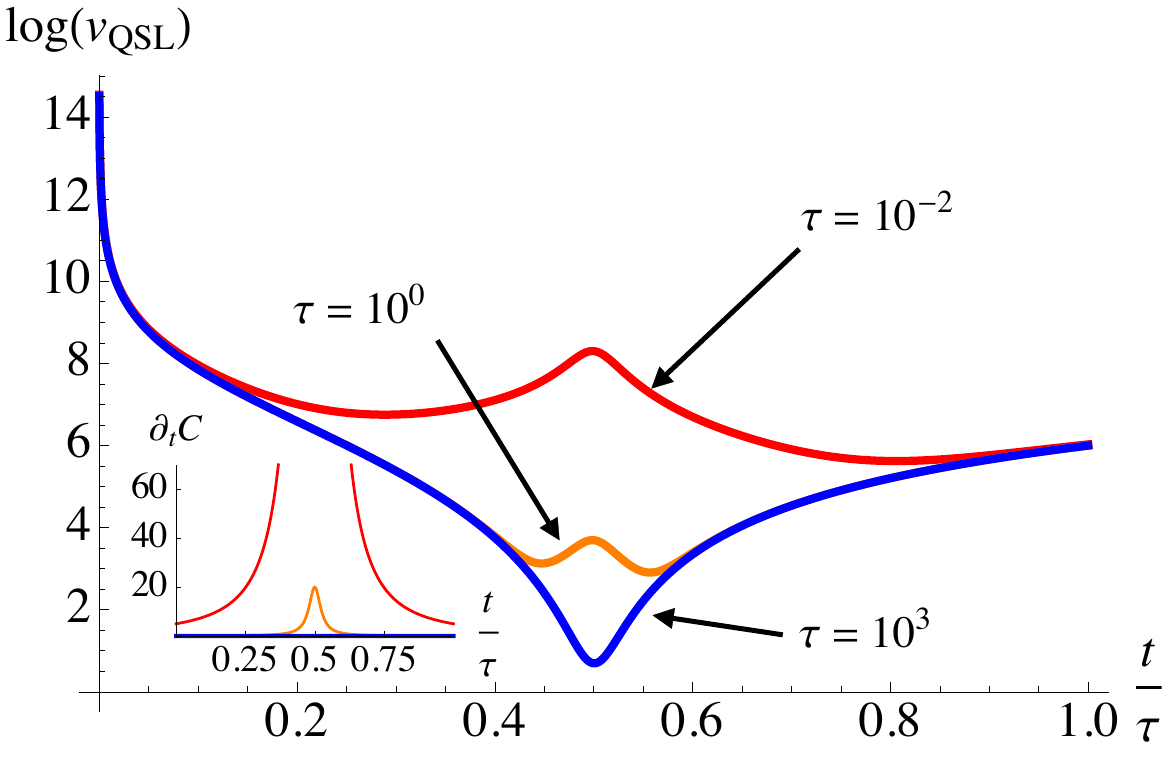}
\caption{Maximal quantum speed $\log(v_\mrm{QSL})$ for the LZ model evolved through the AC using the linear ramp $g(t)=0.2 - 0.4 \tfrac{t}{\tau}$. {\bf (a)} $\Delta=0.001$ and {\bf (b)} $\Delta=0.01$. Both insets show the corresponding instantaneous cost $\partial_t C$.}
\label{fig:LZ}
\end{figure}

In Fig.~\ref{fig:LZ} we examine the role the splitting and total evolution time plays in setting the maximal speed which the system can be driven through the AC using the simple linear ramp $g(t)=g_0 + g_d (t/\tau)$ \footnote{Note that in the unscaled case \eqref{eq14},  $v_\mrm{QSL}$ vanishes for the minimal, zero gap, cf. Eq.~\eqref{eq07}. After rescaling the minimal gap is finite, and hence also the speed remains finite. Therefore, we plot $\log(v_\mrm{QSL})$ to avoid misinterpretation of our findings due to rescaling.}. In panel {\bf (a)} we set $\Delta=0.001$ and consider $\tau=10^3$ (bottom curve), which is close to the adiabatic limit and therefore $\partial_t C \simeq 0$. We observe that the speed steadily decreases as we approach the AC and cusps at $t=0.5\tau$. Comparing to the same evolution time for a larger splitting, $\Delta=0.01$, while the same qualitative behavior is observed we see that the cusp is smoothed out. This has a clear physical interpretation: close to the adiabatic limit we can drive the system at a finite speed far from the AC, however the vanishingly small gap means as we approach the AC we must drive the system extremely slowly, approaching a speed of zero, in order to avoid the excitations that are more likely to occur. Increasing the splitting allows for an increase in the speed at which we can still evolve the system effectively adiabatically. Physically, this is the same behavior that we found for the harmonic oscillator in Fig.~\ref{fig:HO}.

Achieving the same evolution in shorter times requires the use of the counterdiabatic field \eqref{eq15}. In Fig.~\ref{fig:LZ} we see as the system approaches the AC the speed using the counterdiabatic field \emph{increases}. This behavior can again be understood with the help of Fermi's golden rule for time-dependent perturbations \cite{Messiah1966}. The transition probabilities for the `unperturbed' problem are proportional to the time-integrated perturbation. Hence, if the relative magnitude of the perturbation is large, i.e., if the gap is small, transitions can be suppressed if the quantum system is prohibited from lingering at the AC.

A further interesting feature is the clear emergence of a `critical' region, that is delicately dependent on both the splitting and the evolution time. It is clear in both panels that when sufficiently far from the AC, $v_\mrm{QSL}$ is largely independent of evolution time and in these regions the instantaneous cost is close to zero. Approaching the AC requires that either the evolution is accordingly slowed down or a counterdiabatic field is used. This behavior is typical for critical systems, and this is also what is at the core of the Kibble-Zurek mechanism \cite{Francuz2016}. Far away from the critical point the dynamics is essentially adiabatic. However, close to the phase transition the response of the system `freezes out', and the so-called impulse regime emerges. The longer a system lingers in the impulse regime the higher the chances for a transition to occur. Suppressing excitations in the impulse regime, however, is costly, and therefore TQD seeks to rapidly drive the system back into the adiabatic regime.

\paragraph*{Concluding remarks.}
We have achieved three important results: (i) We have rigorously proven the relationship between the cost of TQD and QSL. (ii) We have elucidated the trade-off between speed and cost of a shortcut to adiabaticity. (iii) Finally, by illustrating our general findings with two experimentally relevant systems, we have highlighted the crucial role of the gap for the cost and speed with which a shortcut can be facilitated. In particular, we have found that effectively instantaneous, yet adiabatic dynamics can be achieved at the expense of an infinite cost.

Interestingly, in its original formulation, TQD does not provide any physical intuition as to why it can achieve such fast dynamics. Furthermore, it is reasonable to assume that even using TQD, a small energy gap would require slower driving. Our results show that the shortcut comes from the increased allowed speed of evolution and, quite counterintuitively, TQD encourages faster driving when the energy gap closes. Such an insight could be highly relevant for experimental implementations of TQD~\cite{Bason2012,Zhang2013}.

Our analysis of the LZ model extrapolates to many critical spins systems, such as the Ising and the LMG model. In any system, for which the  gap vanishes as $N \!\! \to \!\! \infty$, our findings for small $\Delta$ qualitatively apply (see also Ref.~\cite{Campbell2016}). Hence, our analysis is particularly important for current efforts in building and improveing quantum computing hardware \cite{Bian2010,Johansson2009,Johnson2011}. Finally, there are two immediate directions for generalizations of our results: open systems \cite{Vacanti2014} and non-Schr\"odinger dynamics \cite{Deffner2016}. While we expect such an intuitive trade-off to persist, a reasonable notion of a thermodynamic cost will have to be found, first.

\acknowledgements{SC acknowledges support from the EU Collaborative Project TherMiQ (Grant Agreement 618074), the Julian Schwinger Foundation (JSF-14-7-0000), and COST Action MP1209 ``Thermodynamics in the quantum regime". SD acknowledges support by the U.S. National Science Foundation under Grant No. CHE-1648973.}

\bibliography{sta_qsl}

\begin{thebibliography}{88}%
\makeatletter
\providecommand \@ifxundefined [1]{%
 \@ifx{#1\undefined}
}%
\providecommand \@ifnum [1]{%
 \ifnum #1\expandafter \@firstoftwo
 \else \expandafter \@secondoftwo
 \fi
}%
\providecommand \@ifx [1]{%
 \ifx #1\expandafter \@firstoftwo
 \else \expandafter \@secondoftwo
 \fi
}%
\providecommand \natexlab [1]{#1}%
\providecommand \enquote  [1]{``#1''}%
\providecommand \bibnamefont  [1]{#1}%
\providecommand \bibfnamefont [1]{#1}%
\providecommand \citenamefont [1]{#1}%
\providecommand \href@noop [0]{\@secondoftwo}%
\providecommand \href [0]{\begingroup \@sanitize@url \@href}%
\providecommand \@href[1]{\@@startlink{#1}\@@href}%
\providecommand \@@href[1]{\endgroup#1\@@endlink}%
\providecommand \@sanitize@url [0]{\catcode `\\12\catcode `\$12\catcode
  `\&12\catcode `\#12\catcode `\^12\catcode `\_12\catcode `\%12\relax}%
\providecommand \@@startlink[1]{}%
\providecommand \@@endlink[0]{}%
\providecommand \url  [0]{\begingroup\@sanitize@url \@url }%
\providecommand \@url [1]{\endgroup\@href {#1}{\urlprefix }}%
\providecommand \urlprefix  [0]{URL }%
\providecommand \Eprint [0]{\href }%
\providecommand \doibase [0]{http://dx.doi.org/}%
\providecommand \selectlanguage [0]{\@gobble}%
\providecommand \bibinfo  [0]{\@secondoftwo}%
\providecommand \bibfield  [0]{\@secondoftwo}%
\providecommand \translation [1]{[#1]}%
\providecommand \BibitemOpen [0]{}%
\providecommand \bibitemStop [0]{}%
\providecommand \bibitemNoStop [0]{.\EOS\space}%
\providecommand \EOS [0]{\spacefactor3000\relax}%
\providecommand \BibitemShut  [1]{\csname bibitem#1\endcsname}%
\let\auto@bib@innerbib\@empty
\bibitem [{\citenamefont {Torrontegui}\ \emph {et~al.}(2013)\citenamefont
  {Torrontegui}, \citenamefont {Ib\'{a}\~{n}ez}, \citenamefont
  {Mart\'{\i}nez-Garaot}, \citenamefont {Modugno}, \citenamefont {del Campo},
  \citenamefont {Gu\'{e}ry-Odelin}, \citenamefont {Ruschhaupt}, \citenamefont
  {Chen},\ and\ \citenamefont {Muga}}]{Torrontegui2013}%
  \BibitemOpen
  \bibfield  {author} {\bibinfo {author} {\bibfnamefont {E.}~\bibnamefont
  {Torrontegui}}, \bibinfo {author} {\bibfnamefont {S.}~\bibnamefont
  {Ib\'{a}\~{n}ez}}, \bibinfo {author} {\bibfnamefont {S.}~\bibnamefont
  {Mart\'{\i}nez-Garaot}}, \bibinfo {author} {\bibfnamefont {M.}~\bibnamefont
  {Modugno}}, \bibinfo {author} {\bibfnamefont {A.}~\bibnamefont {del Campo}},
  \bibinfo {author} {\bibfnamefont {D.}~\bibnamefont {Gu\'{e}ry-Odelin}},
  \bibinfo {author} {\bibfnamefont {A.}~\bibnamefont {Ruschhaupt}}, \bibinfo
  {author} {\bibfnamefont {X.}~\bibnamefont {Chen}}, \ and\ \bibinfo {author}
  {\bibfnamefont {J.~G.}\ \bibnamefont {Muga}},\ }\bibfield  {title} {\enquote
  {\bibinfo {title} {{Shortcuts to Adiabaticity}},}\ }\href {\doibase
  10.1016/B978-0-12-408090-4.00002-5} {\bibfield  {journal} {\bibinfo
  {journal} {Adv. At. Mol. Opt. Phys.}\ }\textbf {\bibinfo {volume} {62}},\
  \bibinfo {pages} {117} (\bibinfo {year} {2013})}\BibitemShut {NoStop}%
\bibitem [{\citenamefont {Campisi}\ \emph {et~al.}(2011)\citenamefont
  {Campisi}, \citenamefont {H\"anggi},\ and\ \citenamefont
  {Talkner}}]{Campisi2011}%
  \BibitemOpen
  \bibfield  {author} {\bibinfo {author} {\bibfnamefont {M.}~\bibnamefont
  {Campisi}}, \bibinfo {author} {\bibfnamefont {P.}~\bibnamefont {H\"anggi}}, \
  and\ \bibinfo {author} {\bibfnamefont {P.}~\bibnamefont {Talkner}},\
  }\bibfield  {title} {\enquote {\bibinfo {title} {\textit{Colloquium} :
  Quantum fluctuation relations: Foundations and applications},}\ }\href
  {\doibase 10.1103/RevModPhys.83.771} {\bibfield  {journal} {\bibinfo
  {journal} {Rev. Mod. Phys.}\ }\textbf {\bibinfo {volume} {83}},\ \bibinfo
  {pages} {771--791} (\bibinfo {year} {2011})}\BibitemShut {NoStop}%
\bibitem [{\citenamefont {Deffner}\ \emph {et~al.}(2016)\citenamefont
  {Deffner}, \citenamefont {Paz},\ and\ \citenamefont
  {Zurek}}]{Deffner2016PRE}%
  \BibitemOpen
  \bibfield  {author} {\bibinfo {author} {\bibfnamefont {S.}~\bibnamefont
  {Deffner}}, \bibinfo {author} {\bibfnamefont {J.~P.}\ \bibnamefont {Paz}}, \
  and\ \bibinfo {author} {\bibfnamefont {W.~H.}\ \bibnamefont {Zurek}},\
  }\bibfield  {title} {\enquote {\bibinfo {title} {Quantum work and the
  thermodynamic cost of quantum measurements},}\ }\href {\doibase
  10.1103/PhysRevE.94.010103} {\bibfield  {journal} {\bibinfo  {journal} {Phys.
  Rev. E}\ }\textbf {\bibinfo {volume} {94}},\ \bibinfo {pages} {010103}
  (\bibinfo {year} {2016})}\BibitemShut {NoStop}%
\bibitem [{\citenamefont {Acconcia}\ and\ \citenamefont
  {Bonan\ifmmode~\mbox{\c{c}}\else \c{c}\fi{}a}(2015)}]{Acconcia2015a}%
  \BibitemOpen
  \bibfield  {author} {\bibinfo {author} {\bibfnamefont {T.~V.}\ \bibnamefont
  {Acconcia}}\ and\ \bibinfo {author} {\bibfnamefont {Marcus V.~S.}\
  \bibnamefont {Bonan\ifmmode~\mbox{\c{c}}\else \c{c}\fi{}a}},\ }\bibfield
  {title} {\enquote {\bibinfo {title} {Degenerate optimal paths in thermally
  isolated systems},}\ }\href {\doibase 10.1103/PhysRevE.91.042141} {\bibfield
  {journal} {\bibinfo  {journal} {Phys. Rev. E}\ }\textbf {\bibinfo {volume}
  {91}},\ \bibinfo {pages} {042141} (\bibinfo {year} {2015})}\BibitemShut
  {NoStop}%
\bibitem [{\citenamefont {Acconcia}\ \emph {et~al.}(2015)\citenamefont
  {Acconcia}, \citenamefont {Bonan\ifmmode~\mbox{\c{c}}\else \c{c}\fi{}a},\
  and\ \citenamefont {Deffner}}]{Acconcia2015}%
  \BibitemOpen
  \bibfield  {author} {\bibinfo {author} {\bibfnamefont {T.~V.}\ \bibnamefont
  {Acconcia}}, \bibinfo {author} {\bibfnamefont {M.~V.~S.}\ \bibnamefont
  {Bonan\ifmmode~\mbox{\c{c}}\else \c{c}\fi{}a}}, \ and\ \bibinfo {author}
  {\bibfnamefont {S.}~\bibnamefont {Deffner}},\ }\bibfield  {title} {\enquote
  {\bibinfo {title} {Shortcuts to adiabaticity from linear response theory},}\
  }\href {\doibase 10.1103/PhysRevE.92.042148} {\bibfield  {journal} {\bibinfo
  {journal} {Phys. Rev. E}\ }\textbf {\bibinfo {volume} {92}},\ \bibinfo
  {pages} {042148} (\bibinfo {year} {2015})}\BibitemShut {NoStop}%
\bibitem [{\citenamefont {{Acconcia}}\ and\ \citenamefont {{Bonan{\c
  c}a}}(2016)}]{Acconcia2016}%
  \BibitemOpen
  \bibfield  {author} {\bibinfo {author} {\bibfnamefont {T.~V.}\ \bibnamefont
  {{Acconcia}}}\ and\ \bibinfo {author} {\bibfnamefont {M.~V.~S.}\ \bibnamefont
  {{Bonan{\c c}a}}},\ }\bibfield  {title} {\enquote {\bibinfo {title}
  {{Microcanonical Szilard engines beyond the quasistatic regime}},}\
  }\href@noop {} {\bibfield  {journal} {\bibinfo  {journal} {ArXiv e-prints}\ }
  (\bibinfo {year} {2016})},\ \Eprint {http://arxiv.org/abs/1606.00944}
  {arXiv:1606.00944 [cond-mat.stat-mech]} \BibitemShut {NoStop}%
\bibitem [{\citenamefont {Chen}\ \emph {et~al.}(2010)\citenamefont {Chen},
  \citenamefont {Ruschhaupt}, \citenamefont {Schmidt}, \citenamefont {del
  Campo}, \citenamefont {Gu\'ery-Odelin},\ and\ \citenamefont
  {Muga}}]{Chen2010}%
  \BibitemOpen
  \bibfield  {author} {\bibinfo {author} {\bibfnamefont {X.}~\bibnamefont
  {Chen}}, \bibinfo {author} {\bibfnamefont {A.}~\bibnamefont {Ruschhaupt}},
  \bibinfo {author} {\bibfnamefont {S.}~\bibnamefont {Schmidt}}, \bibinfo
  {author} {\bibfnamefont {A.}~\bibnamefont {del Campo}}, \bibinfo {author}
  {\bibfnamefont {D.}~\bibnamefont {Gu\'ery-Odelin}}, \ and\ \bibinfo {author}
  {\bibfnamefont {J.~G.}\ \bibnamefont {Muga}},\ }\bibfield  {title} {\enquote
  {\bibinfo {title} {Fast optimal frictionless atom cooling in harmonic traps:
  Shortcut to adiabaticity},}\ }\href {\doibase 10.1103/PhysRevLett.104.063002}
  {\bibfield  {journal} {\bibinfo  {journal} {\PRL}\ }\textbf {\bibinfo
  {volume} {104}},\ \bibinfo {pages} {063002} (\bibinfo {year}
  {2010})}\BibitemShut {NoStop}%
\bibitem [{\citenamefont {del Campo}\ and\ \citenamefont
  {Boshier}(2012)}]{Campo2012}%
  \BibitemOpen
  \bibfield  {author} {\bibinfo {author} {\bibfnamefont {A.}~\bibnamefont {del
  Campo}}\ and\ \bibinfo {author} {\bibfnamefont {M.~G.}\ \bibnamefont
  {Boshier}},\ }\bibfield  {title} {\enquote {\bibinfo {title} {Shortcuts to
  adiabaticity in a time-dependent box},}\ }\href {\doibase 10.1038/srep00648}
  {\bibfield  {journal} {\bibinfo  {journal} {Sci. Rep.}\ }\textbf {\bibinfo
  {volume} {2}},\ \bibinfo {pages} {648} (\bibinfo {year} {2012})}\BibitemShut
  {NoStop}%
\bibitem [{\citenamefont {Masuda}\ and\ \citenamefont
  {Nakamura}(2010)}]{Masuda2010}%
  \BibitemOpen
  \bibfield  {author} {\bibinfo {author} {\bibfnamefont {S.}~\bibnamefont
  {Masuda}}\ and\ \bibinfo {author} {\bibfnamefont {K.}~\bibnamefont
  {Nakamura}},\ }\bibfield  {title} {\enquote {\bibinfo {title} {Fast-forward
  of adiabatic dynamics in quantum mechanics},}\ }\href {\doibase
  http://dx.doi.org/10.1098/rspa.2009.0446} {\bibfield  {journal} {\bibinfo
  {journal} {Proc. R. Soc. A}\ }\textbf {\bibinfo {volume} {466}},\ \bibinfo
  {pages} {1135} (\bibinfo {year} {2010})}\BibitemShut {NoStop}%
\bibitem [{\citenamefont {Masuda}\ and\ \citenamefont
  {Nakamura}(2011)}]{Masuda2011}%
  \BibitemOpen
  \bibfield  {author} {\bibinfo {author} {\bibfnamefont {S.}~\bibnamefont
  {Masuda}}\ and\ \bibinfo {author} {\bibfnamefont {K.}~\bibnamefont
  {Nakamura}},\ }\bibfield  {title} {\enquote {\bibinfo {title} {Acceleration
  of adiabatic quantum dynamics in electromagnetic fields},}\ }\href {\doibase
  http://dx.doi.org/10.1103/PhysRevA.84.043434} {\bibfield  {journal} {\bibinfo
   {journal} {Phys. Rev. A}\ }\textbf {\bibinfo {volume} {84}},\ \bibinfo
  {pages} {043434} (\bibinfo {year} {2011})}\BibitemShut {NoStop}%
\bibitem [{\citenamefont {Torrontegui}\ \emph
  {et~al.}(2012{\natexlab{a}})\citenamefont {Torrontegui}, \citenamefont
  {Mart\'{\i}nez-Garaot}, \citenamefont {Ruschhaupt},\ and\ \citenamefont
  {Muga}}]{Torrontegui2012}%
  \BibitemOpen
  \bibfield  {author} {\bibinfo {author} {\bibfnamefont {E.}~\bibnamefont
  {Torrontegui}}, \bibinfo {author} {\bibfnamefont {S.}~\bibnamefont
  {Mart\'{\i}nez-Garaot}}, \bibinfo {author} {\bibfnamefont {A.}~\bibnamefont
  {Ruschhaupt}}, \ and\ \bibinfo {author} {\bibfnamefont {J.~G.}\ \bibnamefont
  {Muga}},\ }\bibfield  {title} {\enquote {\bibinfo {title} {Shortcuts to
  adiabaticity: Fast-forward approach},}\ }\href {\doibase
  10.1103/PhysRevA.86.013601} {\bibfield  {journal} {\bibinfo  {journal} {Phys.
  Rev. A}\ }\textbf {\bibinfo {volume} {86}},\ \bibinfo {pages} {013601}
  (\bibinfo {year} {2012}{\natexlab{a}})}\BibitemShut {NoStop}%
\bibitem [{\citenamefont {Torrontegui}\ \emph
  {et~al.}(2012{\natexlab{b}})\citenamefont {Torrontegui}, \citenamefont
  {Chen}, \citenamefont {Modugno}, \citenamefont {Schmidt}, \citenamefont
  {Ruschhaupt},\ and\ \citenamefont {Muga}}]{Torrontegui2012a}%
  \BibitemOpen
  \bibfield  {author} {\bibinfo {author} {\bibfnamefont {E.}~\bibnamefont
  {Torrontegui}}, \bibinfo {author} {\bibfnamefont {X.}~\bibnamefont {Chen}},
  \bibinfo {author} {\bibfnamefont {M.}~\bibnamefont {Modugno}}, \bibinfo
  {author} {\bibfnamefont {S.}~\bibnamefont {Schmidt}}, \bibinfo {author}
  {\bibfnamefont {A.}~\bibnamefont {Ruschhaupt}}, \ and\ \bibinfo {author}
  {\bibfnamefont {J.~G.}\ \bibnamefont {Muga}},\ }\bibfield  {title} {\enquote
  {\bibinfo {title} {{Fast transport of Bose–-Einstein condensates}},}\
  }\href {\doibase 10.1088/1367-2630/14/1/013031} {\bibfield  {journal}
  {\bibinfo  {journal} {New J. Phys.}\ }\textbf {\bibinfo {volume} {14}},\
  \bibinfo {pages} {013031} (\bibinfo {year} {2012}{\natexlab{b}})}\BibitemShut
  {NoStop}%
\bibitem [{\citenamefont {Masuda}\ \emph {et~al.}(2014)\citenamefont {Masuda},
  \citenamefont {Nakamura},\ and\ \citenamefont {del Campo}}]{Masuda2014a}%
  \BibitemOpen
  \bibfield  {author} {\bibinfo {author} {\bibfnamefont {S.}~\bibnamefont
  {Masuda}}, \bibinfo {author} {\bibfnamefont {K.}~\bibnamefont {Nakamura}}, \
  and\ \bibinfo {author} {\bibfnamefont {A.}~\bibnamefont {del Campo}},\
  }\bibfield  {title} {\enquote {\bibinfo {title} {High-fidelity rapid
  ground-state loading of an ultracold gas into an optical lattice},}\ }\href
  {\doibase 10.1103/PhysRevLett.113.063003} {\bibfield  {journal} {\bibinfo
  {journal} {Phys. Rev. Lett.}\ }\textbf {\bibinfo {volume} {113}},\ \bibinfo
  {pages} {063003} (\bibinfo {year} {2014})}\BibitemShut {NoStop}%
\bibitem [{\citenamefont {Kiely}\ \emph {et~al.}(2015)\citenamefont {Kiely},
  \citenamefont {McGuinness}, \citenamefont {Muga},\ and\ \citenamefont
  {Ruschhaupt}}]{Kiely2015}%
  \BibitemOpen
  \bibfield  {author} {\bibinfo {author} {\bibfnamefont {A.}~\bibnamefont
  {Kiely}}, \bibinfo {author} {\bibfnamefont {J.~P.~L.}\ \bibnamefont
  {McGuinness}}, \bibinfo {author} {\bibfnamefont {J.~G.}\ \bibnamefont
  {Muga}}, \ and\ \bibinfo {author} {\bibfnamefont {A.}~\bibnamefont
  {Ruschhaupt}},\ }\bibfield  {title} {\enquote {\bibinfo {title} {Fast and
  stable manipulation of a charged particle in a penning trap},}\ }\href
  {http://stacks.iop.org/0953-4075/48/i=7/a=075503} {\bibfield  {journal}
  {\bibinfo  {journal} {J. Phys. B: At. Mol. Opt. Phys.}\ }\textbf {\bibinfo
  {volume} {48}},\ \bibinfo {pages} {075503} (\bibinfo {year}
  {2015})}\BibitemShut {NoStop}%
\bibitem [{\citenamefont {Deffner}(2016)}]{Deffner2016}%
  \BibitemOpen
  \bibfield  {author} {\bibinfo {author} {\bibfnamefont {S.}~\bibnamefont
  {Deffner}},\ }\bibfield  {title} {\enquote {\bibinfo {title} {Shortcuts to
  adiabaticity: suppression of pair production in driven dirac dynamics},}\
  }\href {http://stacks.iop.org/1367-2630/18/i=1/a=012001} {\bibfield
  {journal} {\bibinfo  {journal} {New J. Phys.}\ }\textbf {\bibinfo {volume}
  {18}},\ \bibinfo {pages} {012001} (\bibinfo {year} {2016})}\BibitemShut
  {NoStop}%
\bibitem [{\citenamefont {Demirplak}\ and\ \citenamefont
  {Rice}(2003)}]{Demirplak2003}%
  \BibitemOpen
  \bibfield  {author} {\bibinfo {author} {\bibfnamefont {M.}~\bibnamefont
  {Demirplak}}\ and\ \bibinfo {author} {\bibfnamefont {S.~A.}\ \bibnamefont
  {Rice}},\ }\bibfield  {title} {\enquote {\bibinfo {title} {Adiabatic
  population transfer with control fields},}\ }\href {\doibase
  10.1021/jp030708a} {\bibfield  {journal} {\bibinfo  {journal} {J. Chem. Phys.
  A}\ }\textbf {\bibinfo {volume} {107}},\ \bibinfo {pages} {9937} (\bibinfo
  {year} {2003})}\BibitemShut {NoStop}%
\bibitem [{\citenamefont {Demirplak}\ and\ \citenamefont
  {Rice}(2005)}]{Demirplak2005}%
  \BibitemOpen
  \bibfield  {author} {\bibinfo {author} {\bibfnamefont {M.}~\bibnamefont
  {Demirplak}}\ and\ \bibinfo {author} {\bibfnamefont {S.~A}\ \bibnamefont
  {Rice}},\ }\bibfield  {title} {\enquote {\bibinfo {title} {{Assisted
  adiabatic passage revisited}},}\ }\href {\doibase
  http://dx.doi.org/10.1021/jp040647w} {\bibfield  {journal} {\bibinfo
  {journal} {J. Phys. Chem. B}\ }\textbf {\bibinfo {volume} {109}},\ \bibinfo
  {pages} {6838} (\bibinfo {year} {2005})}\BibitemShut {NoStop}%
\bibitem [{\citenamefont {Berry}(2009)}]{Berry2009}%
  \BibitemOpen
  \bibfield  {author} {\bibinfo {author} {\bibfnamefont {M.}~\bibnamefont
  {Berry}},\ }\bibfield  {title} {\enquote {\bibinfo {title} {Transitionless
  quantum driving},}\ }\href {\doibase 10.1088/1751-8113/42/36/365303}
  {\bibfield  {journal} {\bibinfo  {journal} {J. Phys. A: Math. Theor.}\
  }\textbf {\bibinfo {volume} {42}},\ \bibinfo {pages} {365303} (\bibinfo
  {year} {2009})}\BibitemShut {NoStop}%
\bibitem [{\citenamefont {Deffner}\ \emph {et~al.}(2014)\citenamefont
  {Deffner}, \citenamefont {Jarzynski},\ and\ \citenamefont {del
  Campo}}]{Deffner2014}%
  \BibitemOpen
  \bibfield  {author} {\bibinfo {author} {\bibfnamefont {S.}~\bibnamefont
  {Deffner}}, \bibinfo {author} {\bibfnamefont {C.}~\bibnamefont {Jarzynski}},
  \ and\ \bibinfo {author} {\bibfnamefont {A.}~\bibnamefont {del Campo}},\
  }\bibfield  {title} {\enquote {\bibinfo {title} {{Classical and quantum
  shortcuts to adiabaticity for scale-invariant driving}},}\ }\href {\doibase
  10.1103/PhysRevX.4.021013} {\bibfield  {journal} {\bibinfo  {journal} {Phys.
  Rev. X}\ }\textbf {\bibinfo {volume} {4}},\ \bibinfo {pages} {021013}
  (\bibinfo {year} {2014})}\BibitemShut {NoStop}%
\bibitem [{\citenamefont {{Jarzynski}}(2013)}]{Jarzysnki2013}%
  \BibitemOpen
  \bibfield  {author} {\bibinfo {author} {\bibfnamefont {C.}~\bibnamefont
  {{Jarzynski}}},\ }\bibfield  {title} {\enquote {\bibinfo {title} {{Generating
  shortcuts to adiabaticity in quantum and classical dynamics}},}\ }\href
  {\doibase 10.1103/PhysRevA.88.040101} {\bibfield  {journal} {\bibinfo
  {journal} {\pra}\ }\textbf {\bibinfo {volume} {88}},\ \bibinfo {eid} {040101}
  (\bibinfo {year} {2013})}\BibitemShut {NoStop}%
\bibitem [{\citenamefont {{Patra}}\ and\ \citenamefont
  {{Jarzynski}}(2016)}]{Patra2016}%
  \BibitemOpen
  \bibfield  {author} {\bibinfo {author} {\bibfnamefont {A.}~\bibnamefont
  {{Patra}}}\ and\ \bibinfo {author} {\bibfnamefont {C.}~\bibnamefont
  {{Jarzynski}}},\ }\bibfield  {title} {\enquote {\bibinfo {title} {{Classical
  and quantum shortcuts to adiabaticity in a tilted piston}},}\ }\href@noop {}
  {\bibfield  {journal} {\bibinfo  {journal} {ArXiv e-prints}\ } (\bibinfo
  {year} {2016})},\ \Eprint {http://arxiv.org/abs/1608.08996} {arXiv:1608.08996
  [quant-ph]} \BibitemShut {NoStop}%
\bibitem [{\citenamefont {Khaneja}\ \emph {et~al.}(2005)\citenamefont
  {Khaneja}, \citenamefont {Reiss}, \citenamefont {Kehlet}, \citenamefont
  {Schulte-Herbrüggen},\ and\ \citenamefont {Glaser}}]{Khaneja2005}%
  \BibitemOpen
  \bibfield  {author} {\bibinfo {author} {\bibfnamefont {N.}~\bibnamefont
  {Khaneja}}, \bibinfo {author} {\bibfnamefont {T.}~\bibnamefont {Reiss}},
  \bibinfo {author} {\bibfnamefont {C.}~\bibnamefont {Kehlet}}, \bibinfo
  {author} {\bibfnamefont {T.}~\bibnamefont {Schulte-Herbrüggen}}, \ and\
  \bibinfo {author} {\bibfnamefont {S.~J.}\ \bibnamefont {Glaser}},\ }\bibfield
   {title} {\enquote {\bibinfo {title} {Optimal control of coupled spin
  dynamics: design of {NMR} pulse sequences by gradient ascent algorithms},}\
  }\href {\doibase http://dx.doi.org/10.1016/j.jmr.2004.11.004} {\bibfield
  {journal} {\bibinfo  {journal} {J. Mag. Res.}\ }\textbf {\bibinfo {volume}
  {172}},\ \bibinfo {pages} {296} (\bibinfo {year} {2005})}\BibitemShut
  {NoStop}%
\bibitem [{\citenamefont {Chen}\ \emph {et~al.}(2011)\citenamefont {Chen},
  \citenamefont {Torrontegui}, \citenamefont {Stefanatos}, \citenamefont {Li},\
  and\ \citenamefont {Muga}}]{Chen2011a}%
  \BibitemOpen
  \bibfield  {author} {\bibinfo {author} {\bibfnamefont {X.}~\bibnamefont
  {Chen}}, \bibinfo {author} {\bibfnamefont {E.}~\bibnamefont {Torrontegui}},
  \bibinfo {author} {\bibfnamefont {D.}~\bibnamefont {Stefanatos}}, \bibinfo
  {author} {\bibfnamefont {J.}~\bibnamefont {Li}}, \ and\ \bibinfo {author}
  {\bibfnamefont {J.~G.}\ \bibnamefont {Muga}},\ }\bibfield  {title} {\enquote
  {\bibinfo {title} {Optimal trajectories for efficient atomic transport
  without final excitation},}\ }\href {\doibase
  http://dx.doi.org/10.1103/PhysRevA.84.043415} {\bibfield  {journal} {\bibinfo
   {journal} {Phys. Rev. A}\ }\textbf {\bibinfo {volume} {84}},\ \bibinfo
  {pages} {043415} (\bibinfo {year} {2011})}\BibitemShut {NoStop}%
\bibitem [{\citenamefont {Stefanatos}(2013)}]{Stefanatos2013}%
  \BibitemOpen
  \bibfield  {author} {\bibinfo {author} {\bibfnamefont {D.}~\bibnamefont
  {Stefanatos}},\ }\bibfield  {title} {\enquote {\bibinfo {title} {Optimal
  shortcuts to adiabaticity for a quantum piston},}\ }\href {\doibase
  http://dx.doi.org/10.1016/j.automatica.2013.07.020} {\bibfield  {journal}
  {\bibinfo  {journal} {Automatica}\ }\textbf {\bibinfo {volume} {49}},\
  \bibinfo {pages} {3079} (\bibinfo {year} {2013})}\BibitemShut {NoStop}%
\bibitem [{\citenamefont {Campbell}\ \emph {et~al.}(2015)\citenamefont
  {Campbell}, \citenamefont {{De Chiara}}, \citenamefont {Paternostro},
  \citenamefont {Palma},\ and\ \citenamefont {Fazio}}]{Campbell2014}%
  \BibitemOpen
  \bibfield  {author} {\bibinfo {author} {\bibfnamefont {S.}~\bibnamefont
  {Campbell}}, \bibinfo {author} {\bibfnamefont {G.}~\bibnamefont {{De
  Chiara}}}, \bibinfo {author} {\bibfnamefont {M.}~\bibnamefont {Paternostro}},
  \bibinfo {author} {\bibfnamefont {G.~M.}\ \bibnamefont {Palma}}, \ and\
  \bibinfo {author} {\bibfnamefont {R.}~\bibnamefont {Fazio}},\ }\bibfield
  {title} {\enquote {\bibinfo {title} {{Shortcut to Adiabaticity in the
  Lipkin-Meshkov-Glick Model}},}\ }\href {\doibase
  10.1103/PhysRevLett.114.177206} {\bibfield  {journal} {\bibinfo  {journal}
  {Phys. Rev. Lett.}\ }\textbf {\bibinfo {volume} {114}},\ \bibinfo {pages}
  {177206} (\bibinfo {year} {2015})}\BibitemShut {NoStop}%
\bibitem [{\citenamefont {Wu}\ \emph {et~al.}(2015)\citenamefont {Wu},
  \citenamefont {Nanduri},\ and\ \citenamefont {Rabitz}}]{Wu2015}%
  \BibitemOpen
  \bibfield  {author} {\bibinfo {author} {\bibfnamefont {N.}~\bibnamefont
  {Wu}}, \bibinfo {author} {\bibfnamefont {A.}~\bibnamefont {Nanduri}}, \ and\
  \bibinfo {author} {\bibfnamefont {H.}~\bibnamefont {Rabitz}},\ }\bibfield
  {title} {\enquote {\bibinfo {title} {Optimal suppression of defect generation
  during a passage across a quantum critical point},}\ }\href {\doibase
  10.1103/PhysRevB.91.041115} {\bibfield  {journal} {\bibinfo  {journal} {Phys.
  Rev. B}\ }\textbf {\bibinfo {volume} {91}},\ \bibinfo {pages} {041115}
  (\bibinfo {year} {2015})}\BibitemShut {NoStop}%
\bibitem [{\citenamefont {Saberi}\ \emph {et~al.}(2014)\citenamefont {Saberi},
  \citenamefont {Opatrn\'y}, \citenamefont {M\o{}lmer},\ and\ \citenamefont
  {del Campo}}]{Saberi2014}%
  \BibitemOpen
  \bibfield  {author} {\bibinfo {author} {\bibfnamefont {H.}~\bibnamefont
  {Saberi}}, \bibinfo {author} {\bibfnamefont {T.}~\bibnamefont {Opatrn\'y}},
  \bibinfo {author} {\bibfnamefont {K.}~\bibnamefont {M\o{}lmer}}, \ and\
  \bibinfo {author} {\bibfnamefont {A.}~\bibnamefont {del Campo}},\ }\bibfield
  {title} {\enquote {\bibinfo {title} {Adiabatic tracking of quantum many-body
  dynamics},}\ }\href {\doibase 10.1103/PhysRevA.90.060301} {\bibfield
  {journal} {\bibinfo  {journal} {Phys. Rev. A}\ }\textbf {\bibinfo {volume}
  {90}},\ \bibinfo {pages} {060301} (\bibinfo {year} {2014})}\BibitemShut
  {NoStop}%
\bibitem [{\citenamefont {{Sels}}\ and\ \citenamefont
  {{Polkovnikov}}(2016)}]{PolkovnikovArXiv}%
  \BibitemOpen
  \bibfield  {author} {\bibinfo {author} {\bibfnamefont {D.}~\bibnamefont
  {{Sels}}}\ and\ \bibinfo {author} {\bibfnamefont {A.}~\bibnamefont
  {{Polkovnikov}}},\ }\bibfield  {title} {\enquote {\bibinfo {title}
  {{Minimizing irreversible losses in quantum systems by local counter-diabatic
  driving}},}\ }\href@noop {} {\bibfield  {journal} {\bibinfo  {journal} {ArXiv
  e-prints}\ } (\bibinfo {year} {2016})},\ \Eprint
  {http://arxiv.org/abs/1607.05687} {arXiv:1607.05687 [quant-ph]} \BibitemShut
  {NoStop}%
\bibitem [{\citenamefont {Xiao}\ and\ \citenamefont {Gong}(2014)}]{Xiao2014}%
  \BibitemOpen
  \bibfield  {author} {\bibinfo {author} {\bibfnamefont {G.}~\bibnamefont
  {Xiao}}\ and\ \bibinfo {author} {\bibfnamefont {J.}~\bibnamefont {Gong}},\
  }\bibfield  {title} {\enquote {\bibinfo {title} {{Suppression of work
  fluctuations by optimal control: An approach based on Jarzynski's
  equality}},}\ }\href {\doibase 10.1103/PhysRevE.90.052132} {\bibfield
  {journal} {\bibinfo  {journal} {Phys. Rev. E}\ }\textbf {\bibinfo {volume}
  {90}},\ \bibinfo {pages} {052132} (\bibinfo {year} {2014})}\BibitemShut
  {NoStop}%
\bibitem [{\citenamefont {{Masuda}}\ and\ \citenamefont
  {{Rice}}(2015)}]{Masuda2014}%
  \BibitemOpen
  \bibfield  {author} {\bibinfo {author} {\bibfnamefont {S.}~\bibnamefont
  {{Masuda}}}\ and\ \bibinfo {author} {\bibfnamefont {S.~A.}\ \bibnamefont
  {{Rice}}},\ }\bibfield  {title} {\enquote {\bibinfo {title} {{Fast-Forward
  Assisted STIRAP}},}\ }\href {\doibase 10.1021/acs.jpca.5b00525} {\bibfield
  {journal} {\bibinfo  {journal} {J. Phys. Chem. A}\ }\textbf {\bibinfo
  {volume} {119}},\ \bibinfo {pages} {3479} (\bibinfo {year}
  {2015})}\BibitemShut {NoStop}%
\bibitem [{\citenamefont {Torrontegui}\ \emph {et~al.}(2014)\citenamefont
  {Torrontegui}, \citenamefont {Mart\'{\i}nez-Garaot},\ and\ \citenamefont
  {Muga}}]{Torrontegui2014}%
  \BibitemOpen
  \bibfield  {author} {\bibinfo {author} {\bibfnamefont {E.}~\bibnamefont
  {Torrontegui}}, \bibinfo {author} {\bibfnamefont {S.}~\bibnamefont
  {Mart\'{\i}nez-Garaot}}, \ and\ \bibinfo {author} {\bibfnamefont {J.~G.}\
  \bibnamefont {Muga}},\ }\bibfield  {title} {\enquote {\bibinfo {title}
  {Hamiltonian engineering via invariants and dynamical algebra},}\ }\href
  {\doibase 10.1103/PhysRevA.89.043408} {\bibfield  {journal} {\bibinfo
  {journal} {Phys. Rev. A}\ }\textbf {\bibinfo {volume} {89}},\ \bibinfo
  {pages} {043408} (\bibinfo {year} {2014})}\BibitemShut {NoStop}%
\bibitem [{\citenamefont {Mart\'{\i}nez-Garaot}\ \emph
  {et~al.}(2015)\citenamefont {Mart\'{\i}nez-Garaot}, \citenamefont
  {Ruschhaupt}, \citenamefont {Gillet}, \citenamefont {Busch},\ and\
  \citenamefont {Muga}}]{Garaot2015}%
  \BibitemOpen
  \bibfield  {author} {\bibinfo {author} {\bibfnamefont {S.}~\bibnamefont
  {Mart\'{\i}nez-Garaot}}, \bibinfo {author} {\bibfnamefont {A.}~\bibnamefont
  {Ruschhaupt}}, \bibinfo {author} {\bibfnamefont {J.}~\bibnamefont {Gillet}},
  \bibinfo {author} {\bibfnamefont {T.}~\bibnamefont {Busch}}, \ and\ \bibinfo
  {author} {\bibfnamefont {J.~G.}\ \bibnamefont {Muga}},\ }\bibfield  {title}
  {\enquote {\bibinfo {title} {Fast quasiadiabatic dynamics},}\ }\href
  {\doibase 10.1103/PhysRevA.92.043406} {\bibfield  {journal} {\bibinfo
  {journal} {Phys. Rev. A}\ }\textbf {\bibinfo {volume} {92}},\ \bibinfo
  {pages} {043406} (\bibinfo {year} {2015})}\BibitemShut {NoStop}%
\bibitem [{\citenamefont {Campo}\ \emph {et~al.}(2014)\citenamefont {Campo},
  \citenamefont {Goold},\ and\ \citenamefont {Paternostro}}]{delCampo2013}%
  \BibitemOpen
  \bibfield  {author} {\bibinfo {author} {\bibfnamefont {A.~del}\ \bibnamefont
  {Campo}}, \bibinfo {author} {\bibfnamefont {J.}~\bibnamefont {Goold}}, \ and\
  \bibinfo {author} {\bibfnamefont {M.}~\bibnamefont {Paternostro}},\
  }\bibfield  {title} {\enquote {\bibinfo {title} {More bang for your buck:
  Super-adiabatic quantum engines},}\ }\href
  {http://dx.doi.org/10.1038/srep06208} {\bibfield  {journal} {\bibinfo
  {journal} {Sci. Rep.}\ }\textbf {\bibinfo {volume} {4}},\ \bibinfo {pages}
  {6208} (\bibinfo {year} {2014})}\BibitemShut {NoStop}%
\bibitem [{\citenamefont {{Zheng}}\ \emph {et~al.}(2016)\citenamefont
  {{Zheng}}, \citenamefont {{Campbell}}, \citenamefont {{De Chiara}},\ and\
  \citenamefont {{Poletti}}}]{Zheng2015}%
  \BibitemOpen
  \bibfield  {author} {\bibinfo {author} {\bibfnamefont {Y.}~\bibnamefont
  {{Zheng}}}, \bibinfo {author} {\bibfnamefont {S.}~\bibnamefont {{Campbell}}},
  \bibinfo {author} {\bibfnamefont {G.}~\bibnamefont {{De Chiara}}}, \ and\
  \bibinfo {author} {\bibfnamefont {D.}~\bibnamefont {{Poletti}}},\ }\bibfield
  {title} {\enquote {\bibinfo {title} {{Cost of counterdiabatic driving and
  work output}},}\ }\href {\doibase 10.1103/PhysRevA.94.042132} {\bibfield
  {journal} {\bibinfo  {journal} {\pra}\ }\textbf {\bibinfo {volume} {94}},\
  \bibinfo {eid} {042132} (\bibinfo {year} {2016})}\BibitemShut {NoStop}%
\bibitem [{\citenamefont {Bekenstein}(1981)}]{Bekenstein1981}%
  \BibitemOpen
  \bibfield  {author} {\bibinfo {author} {\bibfnamefont {J.~D.}\ \bibnamefont
  {Bekenstein}},\ }\bibfield  {title} {\enquote {\bibinfo {title} {{Energy Cost
  of Information Transfer}},}\ }\href {\doibase 10.1103/PhysRevLett.46.623}
  {\bibfield  {journal} {\bibinfo  {journal} {Phys. Rev. Lett.}\ }\textbf
  {\bibinfo {volume} {46}},\ \bibinfo {pages} {623--626} (\bibinfo {year}
  {1981})}\BibitemShut {NoStop}%
\bibitem [{\citenamefont {Lloyd}(2000)}]{lloyd00}%
  \BibitemOpen
  \bibfield  {author} {\bibinfo {author} {\bibfnamefont {S.}~\bibnamefont
  {Lloyd}},\ }\bibfield  {title} {\enquote {\bibinfo {title} {Ultimate physical
  limits to computation},}\ }\href {\doibase 10.1038/35023282} {\bibfield
  {journal} {\bibinfo  {journal} {Nature}\ }\textbf {\bibinfo {volume} {406}},\
  \bibinfo {pages} {1047} (\bibinfo {year} {2000})}\BibitemShut {NoStop}%
\bibitem [{\citenamefont {Deffner}\ and\ \citenamefont
  {Lutz}(2010)}]{Deffner2010}%
  \BibitemOpen
  \bibfield  {author} {\bibinfo {author} {\bibfnamefont {S.}~\bibnamefont
  {Deffner}}\ and\ \bibinfo {author} {\bibfnamefont {E.}~\bibnamefont {Lutz}},\
  }\bibfield  {title} {\enquote {\bibinfo {title} {{Generalized Clausius
  inequality for nonequilibrium quantum processes}},}\ }\href {\doibase
  10.1103/PhysRevLett.105.170402} {\bibfield  {journal} {\bibinfo  {journal}
  {Phys. Rev. Lett.}\ }\textbf {\bibinfo {volume} {105}},\ \bibinfo {pages}
  {170402} (\bibinfo {year} {2010})}\BibitemShut {NoStop}%
\bibitem [{\citenamefont {Giovannetti}\ \emph {et~al.}(2011)\citenamefont
  {Giovannetti}, \citenamefont {Lloyd},\ and\ \citenamefont
  {Maccone}}]{Giovannetti2011}%
  \BibitemOpen
  \bibfield  {author} {\bibinfo {author} {\bibfnamefont {V.}~\bibnamefont
  {Giovannetti}}, \bibinfo {author} {\bibfnamefont {S.}~\bibnamefont {Lloyd}},
  \ and\ \bibinfo {author} {\bibfnamefont {L.}~\bibnamefont {Maccone}},\
  }\bibfield  {title} {\enquote {\bibinfo {title} {{Advances in quantum
  metrology}},}\ }\href {\doibase 10.1038/nphoton.2011.35} {\bibfield
  {journal} {\bibinfo  {journal} {Nat. Photonics}\ }\textbf {\bibinfo {volume}
  {5}},\ \bibinfo {pages} {222--229} (\bibinfo {year} {2011})}\BibitemShut
  {NoStop}%
\bibitem [{\citenamefont {Mandelstam}\ and\ \citenamefont
  {Tamm}(1945)}]{mandelstam45}%
  \BibitemOpen
  \bibfield  {author} {\bibinfo {author} {\bibfnamefont {L.}~\bibnamefont
  {Mandelstam}}\ and\ \bibinfo {author} {\bibfnamefont {I.}~\bibnamefont
  {Tamm}},\ }\bibfield  {title} {\enquote {\bibinfo {title} {The uncertainty
  relation between energy and time in nonrelativistic quantum mechanics},}\
  }\href@noop {} {\bibfield  {journal} {\bibinfo  {journal} {J. Phys.}\
  }\textbf {\bibinfo {volume} {9}},\ \bibinfo {pages} {249} (\bibinfo {year}
  {1945})}\BibitemShut {NoStop}%
\bibitem [{\citenamefont {Bhattacharyya}(1983)}]{Bhattacharyya1983}%
  \BibitemOpen
  \bibfield  {author} {\bibinfo {author} {\bibfnamefont {K}~\bibnamefont
  {Bhattacharyya}},\ }\bibfield  {title} {\enquote {\bibinfo {title} {Quantum
  decay and the mandelstam-tamm-energy inequality},}\ }\href
  {http://stacks.iop.org/0305-4470/16/i=13/a=021} {\bibfield  {journal}
  {\bibinfo  {journal} {J. Phys. A: Math. Gen.}\ }\textbf {\bibinfo {volume}
  {16}},\ \bibinfo {pages} {2993} (\bibinfo {year} {1983})}\BibitemShut
  {NoStop}%
\bibitem [{\citenamefont {Pfeifer}(1993)}]{Pfeifer1993}%
  \BibitemOpen
  \bibfield  {author} {\bibinfo {author} {\bibfnamefont {P.}~\bibnamefont
  {Pfeifer}},\ }\bibfield  {title} {\enquote {\bibinfo {title} {How fast can a
  quantum state change with time?}}\ }\href {\doibase
  10.1103/PhysRevLett.70.3365} {\bibfield  {journal} {\bibinfo  {journal}
  {Phys. Rev. Lett.}\ }\textbf {\bibinfo {volume} {70}},\ \bibinfo {pages}
  {3365} (\bibinfo {year} {1993})}\BibitemShut {NoStop}%
\bibitem [{\citenamefont {Margolus}\ and\ \citenamefont
  {Levitin}(1998)}]{margolus98}%
  \BibitemOpen
  \bibfield  {author} {\bibinfo {author} {\bibfnamefont {N.}~\bibnamefont
  {Margolus}}\ and\ \bibinfo {author} {\bibfnamefont {L.~B.}\ \bibnamefont
  {Levitin}},\ }\bibfield  {title} {\enquote {\bibinfo {title} {The maximum
  speed of dynamical evolution},}\ }\href {\doibase
  10.1016/S0167-2789(98)00054-2} {\bibfield  {journal} {\bibinfo  {journal}
  {Phys. D}\ }\textbf {\bibinfo {volume} {120}},\ \bibinfo {pages} {188}
  (\bibinfo {year} {1998})}\BibitemShut {NoStop}%
\bibitem [{\citenamefont {{Giovannetti}}\ \emph {et~al.}(2004)\citenamefont
  {{Giovannetti}}, \citenamefont {{Lloyd}},\ and\ \citenamefont
  {{Maccone}}}]{Giovannetti2004}%
  \BibitemOpen
  \bibfield  {author} {\bibinfo {author} {\bibfnamefont {V.}~\bibnamefont
  {{Giovannetti}}}, \bibinfo {author} {\bibfnamefont {S.}~\bibnamefont
  {{Lloyd}}}, \ and\ \bibinfo {author} {\bibfnamefont {L.}~\bibnamefont
  {{Maccone}}},\ }\bibfield  {title} {\enquote {\bibinfo {title} {{The speed
  limit of quantum unitary evolution}},}\ }\href {\doibase
  10.1088/1464-4266/6/8/028} {\bibfield  {journal} {\bibinfo  {journal} {J.
  Opt. B}\ }\textbf {\bibinfo {volume} {6}},\ \bibinfo {pages} {807} (\bibinfo
  {year} {2004})}\BibitemShut {NoStop}%
\bibitem [{\citenamefont {Barnes}(2013)}]{Barnes2013a}%
  \BibitemOpen
  \bibfield  {author} {\bibinfo {author} {\bibfnamefont {E.}~\bibnamefont
  {Barnes}},\ }\bibfield  {title} {\enquote {\bibinfo {title} {{Analytically
  solvable two-level quantum systems and Landau-Zener interferometry}},}\
  }\href {\doibase 10.1103/PhysRevA.88.013818} {\bibfield  {journal} {\bibinfo
  {journal} {Phys. Rev. A}\ }\textbf {\bibinfo {volume} {88}},\ \bibinfo
  {pages} {013818} (\bibinfo {year} {2013})}\BibitemShut {NoStop}%
\bibitem [{\citenamefont {{Poggi}}\ \emph {et~al.}(2013)\citenamefont
  {{Poggi}}, \citenamefont {{Lombardo}},\ and\ \citenamefont
  {{Wisniacki}}}]{Poggi2013}%
  \BibitemOpen
  \bibfield  {author} {\bibinfo {author} {\bibfnamefont {P.~M.}\ \bibnamefont
  {{Poggi}}}, \bibinfo {author} {\bibfnamefont {F.~C.}\ \bibnamefont
  {{Lombardo}}}, \ and\ \bibinfo {author} {\bibfnamefont {D.~A.}\ \bibnamefont
  {{Wisniacki}}},\ }\bibfield  {title} {\enquote {\bibinfo {title} {{Quantum
  speed limit and optimal evolution time in a two-level system}},}\ }\href
  {\doibase 10.1209/0295-5075/104/40005} {\bibfield  {journal} {\bibinfo
  {journal} {EPL (Europhysics Letters)}\ }\textbf {\bibinfo {volume} {104}},\
  \bibinfo {eid} {40005} (\bibinfo {year} {2013})}\BibitemShut {NoStop}%
\bibitem [{\citenamefont {Hegerfeldt}(2013)}]{Hegerfeldt2013}%
  \BibitemOpen
  \bibfield  {author} {\bibinfo {author} {\bibfnamefont {G.~C.}\ \bibnamefont
  {Hegerfeldt}},\ }\bibfield  {title} {\enquote {\bibinfo {title} {{Driving at
  the quantum speed limit: Optimal control of a two-level system}},}\ }\href
  {\doibase 10.1103/PhysRevLett.111.260501} {\bibfield  {journal} {\bibinfo
  {journal} {Phys. Rev. Lett.}\ }\textbf {\bibinfo {volume} {111}},\ \bibinfo
  {pages} {260501} (\bibinfo {year} {2013})}\BibitemShut {NoStop}%
\bibitem [{\citenamefont {{Andersson}}\ and\ \citenamefont
  {{Heydari}}(2014)}]{Andersson2014}%
  \BibitemOpen
  \bibfield  {author} {\bibinfo {author} {\bibfnamefont {O.}~\bibnamefont
  {{Andersson}}}\ and\ \bibinfo {author} {\bibfnamefont {H.}~\bibnamefont
  {{Heydari}}},\ }\bibfield  {title} {\enquote {\bibinfo {title} {{Quantum
  speed limits and optimal Hamiltonians for driven systems in mixed states}},}\
  }\href {\doibase 10.1088/1751-8113/47/21/215301} {\bibfield  {journal}
  {\bibinfo  {journal} {J. Phys. A}\ }\textbf {\bibinfo {volume} {47}},\
  \bibinfo {eid} {215301} (\bibinfo {year} {2014})}\BibitemShut {NoStop}%
\bibitem [{\citenamefont {{Deffner}}\ and\ \citenamefont
  {{Lutz}}(2013)}]{Deffner2013}%
  \BibitemOpen
  \bibfield  {author} {\bibinfo {author} {\bibfnamefont {S.}~\bibnamefont
  {{Deffner}}}\ and\ \bibinfo {author} {\bibfnamefont {E.}~\bibnamefont
  {{Lutz}}},\ }\bibfield  {title} {\enquote {\bibinfo {title} {{Energy-time
  uncertainty relation for driven quantum systems}},}\ }\href {\doibase
  10.1088/1751-8113/46/33/335302} {\bibfield  {journal} {\bibinfo  {journal}
  {J. Phys. A}\ }\textbf {\bibinfo {volume} {46}},\ \bibinfo {eid} {335302}
  (\bibinfo {year} {2013})}\BibitemShut {NoStop}%
\bibitem [{\citenamefont {{Richerme}}\ \emph {et~al.}(2014)\citenamefont
  {{Richerme}}, \citenamefont {{Gong}}, \citenamefont {{Lee}}, \citenamefont
  {{Senko}}, \citenamefont {{Smith}}, \citenamefont {{Foss-Feig}},
  \citenamefont {{Michalakis}}, \citenamefont {{Gorshkov}},\ and\ \citenamefont
  {{Monroe}}}]{richerme14}%
  \BibitemOpen
  \bibfield  {author} {\bibinfo {author} {\bibfnamefont {P.}~\bibnamefont
  {{Richerme}}}, \bibinfo {author} {\bibfnamefont {Z.-X.}\ \bibnamefont
  {{Gong}}}, \bibinfo {author} {\bibfnamefont {A.}~\bibnamefont {{Lee}}},
  \bibinfo {author} {\bibfnamefont {C.}~\bibnamefont {{Senko}}}, \bibinfo
  {author} {\bibfnamefont {J.}~\bibnamefont {{Smith}}}, \bibinfo {author}
  {\bibfnamefont {M.}~\bibnamefont {{Foss-Feig}}}, \bibinfo {author}
  {\bibfnamefont {S.}~\bibnamefont {{Michalakis}}}, \bibinfo {author}
  {\bibfnamefont {A.~V.}\ \bibnamefont {{Gorshkov}}}, \ and\ \bibinfo {author}
  {\bibfnamefont {C.}~\bibnamefont {{Monroe}}},\ }\bibfield  {title} {\enquote
  {\bibinfo {title} {{Non-local propagation of correlations in long-range
  interacting quantum systems}},}\ }\href
  {http://www.nature.com/nature/journal/v511/n7508/full/nature13450.html}
  {\bibfield  {journal} {\bibinfo  {journal} {Nature}\ }\textbf {\bibinfo
  {volume} {511}},\ \bibinfo {pages} {198} (\bibinfo {year}
  {2014})}\BibitemShut {NoStop}%
\bibitem [{\citenamefont {Jurcevic}\ \emph {et~al.}(2014)\citenamefont
  {Jurcevic}, \citenamefont {Lanyon}, \citenamefont {Hauke}, \citenamefont
  {Hempel}, \citenamefont {Zoller}, \citenamefont {Blatt},\ and\ \citenamefont
  {Roos}}]{jurcevic14}%
  \BibitemOpen
  \bibfield  {author} {\bibinfo {author} {\bibfnamefont {P.}~\bibnamefont
  {Jurcevic}}, \bibinfo {author} {\bibfnamefont {B.~P.}\ \bibnamefont
  {Lanyon}}, \bibinfo {author} {\bibfnamefont {P.}~\bibnamefont {Hauke}},
  \bibinfo {author} {\bibfnamefont {C.}~\bibnamefont {Hempel}}, \bibinfo
  {author} {\bibfnamefont {P.}~\bibnamefont {Zoller}}, \bibinfo {author}
  {\bibfnamefont {R.}~\bibnamefont {Blatt}}, \ and\ \bibinfo {author}
  {\bibfnamefont {C.~F.}\ \bibnamefont {Roos}},\ }\bibfield  {title} {\enquote
  {\bibinfo {title} {Quasiparticle engineering and entanglement propagation in
  a quantum many-body system},}\ }\href {http://dx.doi.org/10.1038/nature13461}
  {\bibfield  {journal} {\bibinfo  {journal} {Nature}\ }\textbf {\bibinfo
  {volume} {511}},\ \bibinfo {pages} {202} (\bibinfo {year}
  {2014})}\BibitemShut {NoStop}%
\bibitem [{\citenamefont {Deffner}\ and\ \citenamefont
  {Lutz}(2013)}]{Deffner2013PRL}%
  \BibitemOpen
  \bibfield  {author} {\bibinfo {author} {\bibfnamefont {S.}~\bibnamefont
  {Deffner}}\ and\ \bibinfo {author} {\bibfnamefont {E.}~\bibnamefont {Lutz}},\
  }\bibfield  {title} {\enquote {\bibinfo {title} {Quantum speed limit for
  non-{Markovian} dynamics},}\ }\href {\doibase 10.1103/PhysRevLett.111.010402}
  {\bibfield  {journal} {\bibinfo  {journal} {Phys. Rev. Lett.}\ }\textbf
  {\bibinfo {volume} {111}},\ \bibinfo {pages} {010402} (\bibinfo {year}
  {2013})}\BibitemShut {NoStop}%
\bibitem [{\citenamefont {del Campo}\ \emph {et~al.}(2013)\citenamefont {del
  Campo}, \citenamefont {Egusquiza}, \citenamefont {Plenio},\ and\
  \citenamefont {Huelga}}]{delcampo13}%
  \BibitemOpen
  \bibfield  {author} {\bibinfo {author} {\bibfnamefont {A.}~\bibnamefont {del
  Campo}}, \bibinfo {author} {\bibfnamefont {I.~L.}\ \bibnamefont {Egusquiza}},
  \bibinfo {author} {\bibfnamefont {M.~B.}\ \bibnamefont {Plenio}}, \ and\
  \bibinfo {author} {\bibfnamefont {S.~F.}\ \bibnamefont {Huelga}},\ }\bibfield
   {title} {\enquote {\bibinfo {title} {Quantum speed limits in open system
  dynamics},}\ }\href {\doibase 10.1103/PhysRevLett.110.050403} {\bibfield
  {journal} {\bibinfo  {journal} {Phys. Rev. Lett.}\ }\textbf {\bibinfo
  {volume} {110}},\ \bibinfo {pages} {050403} (\bibinfo {year}
  {2013})}\BibitemShut {NoStop}%
\bibitem [{\citenamefont {Taddei}\ \emph {et~al.}(2013)\citenamefont {Taddei},
  \citenamefont {Escher}, \citenamefont {Davidovich},\ and\ \citenamefont
  {de~Matos~Filho}}]{taddei13}%
  \BibitemOpen
  \bibfield  {author} {\bibinfo {author} {\bibfnamefont {M.~M.}\ \bibnamefont
  {Taddei}}, \bibinfo {author} {\bibfnamefont {B.~M.}\ \bibnamefont {Escher}},
  \bibinfo {author} {\bibfnamefont {L.}~\bibnamefont {Davidovich}}, \ and\
  \bibinfo {author} {\bibfnamefont {R.~L.}\ \bibnamefont {de~Matos~Filho}},\
  }\bibfield  {title} {\enquote {\bibinfo {title} {Quantum speed limit for
  physical processes},}\ }\href {\doibase 10.1103/PhysRevLett.110.050402}
  {\bibfield  {journal} {\bibinfo  {journal} {Phys. Rev. Lett.}\ }\textbf
  {\bibinfo {volume} {110}},\ \bibinfo {pages} {050402} (\bibinfo {year}
  {2013})}\BibitemShut {NoStop}%
\bibitem [{\citenamefont {{Deffner}}(2014)}]{deffner14}%
  \BibitemOpen
  \bibfield  {author} {\bibinfo {author} {\bibfnamefont {S.}~\bibnamefont
  {{Deffner}}},\ }\bibfield  {title} {\enquote {\bibinfo {title} {{Optimal
  control of a qubit in an optical cavity}},}\ }\href {\doibase
  10.1088/0953-4075/47/14/145502} {\bibfield  {journal} {\bibinfo  {journal}
  {J. Phys. B}\ }\textbf {\bibinfo {volume} {47}},\ \bibinfo {eid} {145502}
  (\bibinfo {year} {2014})}\BibitemShut {NoStop}%
\bibitem [{\citenamefont {{Zhang}}\ \emph {et~al.}(2014)\citenamefont
  {{Zhang}}, \citenamefont {{Han}}, \citenamefont {{Xia}}, \citenamefont
  {{Cao}},\ and\ \citenamefont {{Fan}}}]{Zhang2014}%
  \BibitemOpen
  \bibfield  {author} {\bibinfo {author} {\bibfnamefont {Y.-J.}\ \bibnamefont
  {{Zhang}}}, \bibinfo {author} {\bibfnamefont {W.}~\bibnamefont {{Han}}},
  \bibinfo {author} {\bibfnamefont {Y.-J.}\ \bibnamefont {{Xia}}}, \bibinfo
  {author} {\bibfnamefont {J.-P.}\ \bibnamefont {{Cao}}}, \ and\ \bibinfo
  {author} {\bibfnamefont {H.}~\bibnamefont {{Fan}}},\ }\bibfield  {title}
  {\enquote {\bibinfo {title} {{Quantum speed limit for arbitrary initial
  states}},}\ }\href {\doibase 10.1038/srep04890} {\bibfield  {journal}
  {\bibinfo  {journal} {Sci. Rep.}\ }\textbf {\bibinfo {volume} {4}},\ \bibinfo
  {eid} {4890} (\bibinfo {year} {2014})}\BibitemShut {NoStop}%
\bibitem [{\citenamefont {{Mukherjee}}\ \emph {et~al.}(2013)\citenamefont
  {{Mukherjee}}, \citenamefont {{Carlini}}, \citenamefont {{Mari}},
  \citenamefont {{Caneva}}, \citenamefont {{Montangero}}, \citenamefont
  {{Calarco}}, \citenamefont {{Fazio}},\ and\ \citenamefont
  {{Giovannetti}}}]{Mukherjee2013}%
  \BibitemOpen
  \bibfield  {author} {\bibinfo {author} {\bibfnamefont {V.}~\bibnamefont
  {{Mukherjee}}}, \bibinfo {author} {\bibfnamefont {A.}~\bibnamefont
  {{Carlini}}}, \bibinfo {author} {\bibfnamefont {A.}~\bibnamefont {{Mari}}},
  \bibinfo {author} {\bibfnamefont {T.}~\bibnamefont {{Caneva}}}, \bibinfo
  {author} {\bibfnamefont {S.}~\bibnamefont {{Montangero}}}, \bibinfo {author}
  {\bibfnamefont {T.}~\bibnamefont {{Calarco}}}, \bibinfo {author}
  {\bibfnamefont {R.}~\bibnamefont {{Fazio}}}, \ and\ \bibinfo {author}
  {\bibfnamefont {V.}~\bibnamefont {{Giovannetti}}},\ }\bibfield  {title}
  {\enquote {\bibinfo {title} {{Speeding up and slowing down the relaxation of
  a qubit by optimal control}},}\ }\href {\doibase 10.1103/PhysRevA.88.062326}
  {\bibfield  {journal} {\bibinfo  {journal} {\PRA}\ }\textbf {\bibinfo
  {volume} {88}},\ \bibinfo {eid} {062326} (\bibinfo {year}
  {2013})}\BibitemShut {NoStop}%
\bibitem [{\citenamefont {{Xu}}\ \emph {et~al.}(2014)\citenamefont {{Xu}},
  \citenamefont {{Luo}}, \citenamefont {{Yang}}, \citenamefont {{Liu}},\ and\
  \citenamefont {{Zhu}}}]{Xu2014}%
  \BibitemOpen
  \bibfield  {author} {\bibinfo {author} {\bibfnamefont {Z.-Y.}\ \bibnamefont
  {{Xu}}}, \bibinfo {author} {\bibfnamefont {S.}~\bibnamefont {{Luo}}},
  \bibinfo {author} {\bibfnamefont {W.~L.}\ \bibnamefont {{Yang}}}, \bibinfo
  {author} {\bibfnamefont {C.}~\bibnamefont {{Liu}}}, \ and\ \bibinfo {author}
  {\bibfnamefont {S.}~\bibnamefont {{Zhu}}},\ }\bibfield  {title} {\enquote
  {\bibinfo {title} {Quantum speedup in a memory environment},}\ }\href
  {\doibase 10.1103/PhysRevA.89.012307} {\bibfield  {journal} {\bibinfo
  {journal} {\PRA}\ }\textbf {\bibinfo {volume} {89}},\ \bibinfo {eid} {012307}
  (\bibinfo {year} {2014})}\BibitemShut {NoStop}%
\bibitem [{\citenamefont {{Xu}}\ and\ \citenamefont {{Zhu}}(2014)}]{Xu2014a}%
  \BibitemOpen
  \bibfield  {author} {\bibinfo {author} {\bibfnamefont {Z.-Y.}\ \bibnamefont
  {{Xu}}}\ and\ \bibinfo {author} {\bibfnamefont {S.-Q.}\ \bibnamefont
  {{Zhu}}},\ }\bibfield  {title} {\enquote {\bibinfo {title} {Quantum speed
  limit of a photon under non-{M}arkovian dynamics},}\ }\href {\doibase
  10.1088/0256-307X/31/2/020301} {\bibfield  {journal} {\bibinfo  {journal}
  {Chin. Phys. Lett.}\ }\textbf {\bibinfo {volume} {31}},\ \bibinfo {eid}
  {020301} (\bibinfo {year} {2014})}\BibitemShut {NoStop}%
\bibitem [{\citenamefont {{Cimmarusti}}\ \emph {et~al.}(2015)\citenamefont
  {{Cimmarusti}}, \citenamefont {{Yan}}, \citenamefont {{Patterson}},
  \citenamefont {{Corcos}}, \citenamefont {{Orozco}},\ and\ \citenamefont
  {{Deffner}}}]{Cimmarusti2015}%
  \BibitemOpen
  \bibfield  {author} {\bibinfo {author} {\bibfnamefont {A.~D.}\ \bibnamefont
  {{Cimmarusti}}}, \bibinfo {author} {\bibfnamefont {Z.}~\bibnamefont {{Yan}}},
  \bibinfo {author} {\bibfnamefont {B.~D.}\ \bibnamefont {{Patterson}}},
  \bibinfo {author} {\bibfnamefont {L.~P.}\ \bibnamefont {{Corcos}}}, \bibinfo
  {author} {\bibfnamefont {L.~A.}\ \bibnamefont {{Orozco}}}, \ and\ \bibinfo
  {author} {\bibfnamefont {S.}~\bibnamefont {{Deffner}}},\ }\bibfield  {title}
  {\enquote {\bibinfo {title} {Environment-assisted speed-up of the field
  evolution in cavity quantum electrodynamics},}\ }\href {\doibase
  10.1103/PhysRevLett.114.233602} {\bibfield  {journal} {\bibinfo  {journal}
  {\PRL}\ }\textbf {\bibinfo {volume} {114}},\ \bibinfo {eid} {233602}
  (\bibinfo {year} {2015})}\BibitemShut {NoStop}%
\bibitem [{\citenamefont {Mondal}\ \emph {et~al.}(2016)\citenamefont {Mondal},
  \citenamefont {Datta},\ and\ \citenamefont {Sazim}}]{Mondal2016}%
  \BibitemOpen
  \bibfield  {author} {\bibinfo {author} {\bibfnamefont {D.}~\bibnamefont
  {Mondal}}, \bibinfo {author} {\bibfnamefont {C.}~\bibnamefont {Datta}}, \
  and\ \bibinfo {author} {\bibfnamefont {S.}~\bibnamefont {Sazim}},\ }\bibfield
   {title} {\enquote {\bibinfo {title} {Quantum coherence sets the quantum
  speed limit for mixed states},}\ }\href {\doibase
  http://dx.doi.org/10.1016/j.physleta.2015.12.015} {\bibfield  {journal}
  {\bibinfo  {journal} {Phys. Lett. A}\ }\textbf {\bibinfo {volume} {380}},\
  \bibinfo {pages} {689} (\bibinfo {year} {2016})}\BibitemShut {NoStop}%
\bibitem [{\citenamefont {Mondal}\ and\ \citenamefont
  {Pati}(2016)}]{Mondal2016PLA}%
  \BibitemOpen
  \bibfield  {author} {\bibinfo {author} {\bibfnamefont {D.}~\bibnamefont
  {Mondal}}\ and\ \bibinfo {author} {\bibfnamefont {A.~K.}\ \bibnamefont
  {Pati}},\ }\bibfield  {title} {\enquote {\bibinfo {title} {Quantum speed
  limit for mixed states using an experimentally realizable metric},}\ }\href
  {\doibase http://dx.doi.org/10.1016/j.physleta.2016.02.018} {\bibfield
  {journal} {\bibinfo  {journal} {Phys. Lett. A}\ }\textbf {\bibinfo {volume}
  {380}},\ \bibinfo {pages} {1395} (\bibinfo {year} {2016})}\BibitemShut
  {NoStop}%
\bibitem [{\citenamefont {Deffner}\ and\ \citenamefont
  {Lutz}(2008)}]{Deffner2008}%
  \BibitemOpen
  \bibfield  {author} {\bibinfo {author} {\bibfnamefont {S.}~\bibnamefont
  {Deffner}}\ and\ \bibinfo {author} {\bibfnamefont {E.}~\bibnamefont {Lutz}},\
  }\bibfield  {title} {\enquote {\bibinfo {title} {Nonequilibrium work
  distribution of a quantum harmonic oscillator},}\ }\href {\doibase
  10.1103/PhysRevE.77.021128} {\bibfield  {journal} {\bibinfo  {journal} {Phys.
  Rev. E}\ }\textbf {\bibinfo {volume} {77}},\ \bibinfo {pages} {021128}
  (\bibinfo {year} {2008})}\BibitemShut {NoStop}%
\bibitem [{\citenamefont {{Deffner}}\ \emph {et~al.}(2010)\citenamefont
  {{Deffner}}, \citenamefont {{Abah}},\ and\ \citenamefont
  {{Lutz}}}]{Deffner2010CP}%
  \BibitemOpen
  \bibfield  {author} {\bibinfo {author} {\bibfnamefont {S.}~\bibnamefont
  {{Deffner}}}, \bibinfo {author} {\bibfnamefont {O.}~\bibnamefont {{Abah}}}, \
  and\ \bibinfo {author} {\bibfnamefont {E.}~\bibnamefont {{Lutz}}},\
  }\bibfield  {title} {\enquote {\bibinfo {title} {{Quantum work statistics of
  linear and nonlinear parametric oscillators}},}\ }\href {\doibase
  10.1016/j.chemphys.2010.04.042} {\bibfield  {journal} {\bibinfo  {journal}
  {Chem. Phys.}\ }\textbf {\bibinfo {volume} {375}},\ \bibinfo {pages}
  {200--208} (\bibinfo {year} {2010})}\BibitemShut {NoStop}%
\bibitem [{\citenamefont {{Deffner}}\ and\ \citenamefont
  {{Lutz}}(2013)}]{Deffner2013PRE}%
  \BibitemOpen
  \bibfield  {author} {\bibinfo {author} {\bibfnamefont {S.}~\bibnamefont
  {{Deffner}}}\ and\ \bibinfo {author} {\bibfnamefont {E.}~\bibnamefont
  {{Lutz}}},\ }\bibfield  {title} {\enquote {\bibinfo {title} {{Thermodynamic
  length for far-from-equilibrium quantum systems}},}\ }\href {\doibase
  10.1103/PhysRevE.87.022143} {\bibfield  {journal} {\bibinfo  {journal}
  {\pre}\ }\textbf {\bibinfo {volume} {87}},\ \bibinfo {eid} {022143} (\bibinfo
  {year} {2013})}\BibitemShut {NoStop}%
\bibitem [{\citenamefont {Gong}\ \emph {et~al.}(2014)\citenamefont {Gong},
  \citenamefont {Deffner},\ and\ \citenamefont {Quan}}]{Gong2014}%
  \BibitemOpen
  \bibfield  {author} {\bibinfo {author} {\bibfnamefont {Z.}~\bibnamefont
  {Gong}}, \bibinfo {author} {\bibfnamefont {S.}~\bibnamefont {Deffner}}, \
  and\ \bibinfo {author} {\bibfnamefont {H.~T.}\ \bibnamefont {Quan}},\
  }\bibfield  {title} {\enquote {\bibinfo {title} {Interference of identical
  particles and the quantum work distribution},}\ }\href {\doibase
  10.1103/PhysRevE.90.062121} {\bibfield  {journal} {\bibinfo  {journal} {Phys.
  Rev. E}\ }\textbf {\bibinfo {volume} {90}},\ \bibinfo {pages} {062121}
  (\bibinfo {year} {2014})}\BibitemShut {NoStop}%
\bibitem [{\citenamefont {Huber}\ \emph {et~al.}(2008)\citenamefont {Huber},
  \citenamefont {Schmidt-Kaler}, \citenamefont {Deffner},\ and\ \citenamefont
  {Lutz}}]{Huber2008}%
  \BibitemOpen
  \bibfield  {author} {\bibinfo {author} {\bibfnamefont {G.}~\bibnamefont
  {Huber}}, \bibinfo {author} {\bibfnamefont {F.}~\bibnamefont
  {Schmidt-Kaler}}, \bibinfo {author} {\bibfnamefont {S.}~\bibnamefont
  {Deffner}}, \ and\ \bibinfo {author} {\bibfnamefont {E.}~\bibnamefont
  {Lutz}},\ }\bibfield  {title} {\enquote {\bibinfo {title} {Employing trapped
  cold ions to verify the quantum jarzynski equality},}\ }\href {\doibase
  10.1103/PhysRevLett.101.070403} {\bibfield  {journal} {\bibinfo  {journal}
  {Phys. Rev. Lett.}\ }\textbf {\bibinfo {volume} {101}},\ \bibinfo {pages}
  {070403} (\bibinfo {year} {2008})}\BibitemShut {NoStop}%
\bibitem [{\citenamefont {Abah}\ \emph {et~al.}(2012)\citenamefont {Abah},
  \citenamefont {Ro\ss{}nagel}, \citenamefont {Jacob}, \citenamefont {Deffner},
  \citenamefont {Schmidt-Kaler}, \citenamefont {Singer},\ and\ \citenamefont
  {Lutz}}]{Abah2012}%
  \BibitemOpen
  \bibfield  {author} {\bibinfo {author} {\bibfnamefont {O.}~\bibnamefont
  {Abah}}, \bibinfo {author} {\bibfnamefont {J.}~\bibnamefont {Ro\ss{}nagel}},
  \bibinfo {author} {\bibfnamefont {G.}~\bibnamefont {Jacob}}, \bibinfo
  {author} {\bibfnamefont {S.}~\bibnamefont {Deffner}}, \bibinfo {author}
  {\bibfnamefont {F.}~\bibnamefont {Schmidt-Kaler}}, \bibinfo {author}
  {\bibfnamefont {K.}~\bibnamefont {Singer}}, \ and\ \bibinfo {author}
  {\bibfnamefont {E.}~\bibnamefont {Lutz}},\ }\bibfield  {title} {\enquote
  {\bibinfo {title} {Single-ion heat engine at maximum power},}\ }\href
  {\doibase 10.1103/PhysRevLett.109.203006} {\bibfield  {journal} {\bibinfo
  {journal} {Phys. Rev. Lett.}\ }\textbf {\bibinfo {volume} {109}},\ \bibinfo
  {pages} {203006} (\bibinfo {year} {2012})}\BibitemShut {NoStop}%
\bibitem [{\citenamefont {Ro\ss{}nagel}\ \emph {et~al.}(2014)\citenamefont
  {Ro\ss{}nagel}, \citenamefont {Abah}, \citenamefont {Schmidt-Kaler},
  \citenamefont {Singer},\ and\ \citenamefont {Lutz}}]{Rossnagel2014}%
  \BibitemOpen
  \bibfield  {author} {\bibinfo {author} {\bibfnamefont {J.}~\bibnamefont
  {Ro\ss{}nagel}}, \bibinfo {author} {\bibfnamefont {O.}~\bibnamefont {Abah}},
  \bibinfo {author} {\bibfnamefont {F.}~\bibnamefont {Schmidt-Kaler}}, \bibinfo
  {author} {\bibfnamefont {K.}~\bibnamefont {Singer}}, \ and\ \bibinfo {author}
  {\bibfnamefont {E.}~\bibnamefont {Lutz}},\ }\bibfield  {title} {\enquote
  {\bibinfo {title} {Nanoscale heat engine beyond the carnot limit},}\ }\href
  {\doibase 10.1103/PhysRevLett.112.030602} {\bibfield  {journal} {\bibinfo
  {journal} {Phys. Rev. Lett.}\ }\textbf {\bibinfo {volume} {112}},\ \bibinfo
  {pages} {030602} (\bibinfo {year} {2014})}\BibitemShut {NoStop}%
\bibitem [{\citenamefont {{Ro{\ss}nagel}}\ \emph {et~al.}(2016)\citenamefont
  {{Ro{\ss}nagel}}, \citenamefont {{Dawkins}}, \citenamefont {{Tolazzi}},
  \citenamefont {{Abah}}, \citenamefont {{Lutz}}, \citenamefont
  {{Schmidt-Kaler}},\ and\ \citenamefont {{Singer}}}]{Rossnagel2016}%
  \BibitemOpen
  \bibfield  {author} {\bibinfo {author} {\bibfnamefont {J.}~\bibnamefont
  {{Ro{\ss}nagel}}}, \bibinfo {author} {\bibfnamefont {S.~T.}\ \bibnamefont
  {{Dawkins}}}, \bibinfo {author} {\bibfnamefont {K.~N.}\ \bibnamefont
  {{Tolazzi}}}, \bibinfo {author} {\bibfnamefont {O.}~\bibnamefont {{Abah}}},
  \bibinfo {author} {\bibfnamefont {E.}~\bibnamefont {{Lutz}}}, \bibinfo
  {author} {\bibfnamefont {F.}~\bibnamefont {{Schmidt-Kaler}}}, \ and\ \bibinfo
  {author} {\bibfnamefont {K.}~\bibnamefont {{Singer}}},\ }\bibfield  {title}
  {\enquote {\bibinfo {title} {{A single-atom heat engine}},}\ }\href {\doibase
  10.1126/science.aad6320} {\bibfield  {journal} {\bibinfo  {journal}
  {Science}\ }\textbf {\bibinfo {volume} {352}},\ \bibinfo {pages} {325}
  (\bibinfo {year} {2016})}\BibitemShut {NoStop}%
\bibitem [{\citenamefont {Dziarmaga}(2005)}]{Dziarmaga2005}%
  \BibitemOpen
  \bibfield  {author} {\bibinfo {author} {\bibfnamefont {J.}~\bibnamefont
  {Dziarmaga}},\ }\bibfield  {title} {\enquote {\bibinfo {title} {Dynamics of a
  quantum phase transition: Exact solution of the quantum {I}sing model},}\
  }\href {\doibase 10.1103/PhysRevLett.95.245701} {\bibfield  {journal}
  {\bibinfo  {journal} {Phys. Rev. Lett.}\ }\textbf {\bibinfo {volume} {95}},\
  \bibinfo {pages} {245701} (\bibinfo {year} {2005})}\BibitemShut {NoStop}%
\bibitem [{\citenamefont {del Campo}\ \emph {et~al.}(2012)\citenamefont {del
  Campo}, \citenamefont {Rams},\ and\ \citenamefont {Zurek}}]{delCampo2012LZ}%
  \BibitemOpen
  \bibfield  {author} {\bibinfo {author} {\bibfnamefont {A.}~\bibnamefont {del
  Campo}}, \bibinfo {author} {\bibfnamefont {M.~M.}\ \bibnamefont {Rams}}, \
  and\ \bibinfo {author} {\bibfnamefont {W.~H.}\ \bibnamefont {Zurek}},\
  }\bibfield  {title} {\enquote {\bibinfo {title} {Assisted finite-rate
  adiabatic passage across a quantum critical point: Exact solution for the
  quantum {I}sing model},}\ }\href {\doibase 10.1103/PhysRevLett.109.115703}
  {\bibfield  {journal} {\bibinfo  {journal} {Phys. Rev. Lett.}\ }\textbf
  {\bibinfo {volume} {109}},\ \bibinfo {pages} {115703} (\bibinfo {year}
  {2012})}\BibitemShut {NoStop}%
\bibitem [{\citenamefont {Johansson}\ \emph {et~al.}(2009)\citenamefont
  {Johansson}, \citenamefont {Amin}, \citenamefont {Berkley}, \citenamefont
  {Bunyk}, \citenamefont {Choi}, \citenamefont {Harris}, \citenamefont
  {Johnson}, \citenamefont {Lanting}, \citenamefont {Lloyd},\ and\
  \citenamefont {Rose}}]{Johansson2009}%
  \BibitemOpen
  \bibfield  {author} {\bibinfo {author} {\bibfnamefont {J.}~\bibnamefont
  {Johansson}}, \bibinfo {author} {\bibfnamefont {M.~H.~S.}\ \bibnamefont
  {Amin}}, \bibinfo {author} {\bibfnamefont {A.~J.}\ \bibnamefont {Berkley}},
  \bibinfo {author} {\bibfnamefont {P.}~\bibnamefont {Bunyk}}, \bibinfo
  {author} {\bibfnamefont {V.}~\bibnamefont {Choi}}, \bibinfo {author}
  {\bibfnamefont {R.}~\bibnamefont {Harris}}, \bibinfo {author} {\bibfnamefont
  {M.~W.}\ \bibnamefont {Johnson}}, \bibinfo {author} {\bibfnamefont {T.~M.}\
  \bibnamefont {Lanting}}, \bibinfo {author} {\bibfnamefont {Seth}\
  \bibnamefont {Lloyd}}, \ and\ \bibinfo {author} {\bibfnamefont
  {G.}~\bibnamefont {Rose}},\ }\bibfield  {title} {\enquote {\bibinfo {title}
  {{Landau-Zener transitions in a superconducting flux qubit}},}\ }\href
  {\doibase 10.1103/PhysRevB.80.012507} {\bibfield  {journal} {\bibinfo
  {journal} {Phys. Rev. B}\ }\textbf {\bibinfo {volume} {80}},\ \bibinfo
  {pages} {012507} (\bibinfo {year} {2009})}\BibitemShut {NoStop}%
\bibitem [{\citenamefont {Das}(2009)}]{Das2009}%
  \BibitemOpen
  \bibfield  {author} {\bibinfo {author} {\bibfnamefont {A.}~\bibnamefont
  {Das}},\ }\bibfield  {title} {\enquote {\bibinfo {title} {Non-classical role
  of potential energy in adiabatic quantum annealing},}\ }\href
  {http://stacks.iop.org/1742-6596/143/i=1/a=012001} {\bibfield  {journal}
  {\bibinfo  {journal} {J. Phys.: Conf. Ser.}\ }\textbf {\bibinfo {volume}
  {143}},\ \bibinfo {pages} {012001} (\bibinfo {year} {2009})}\BibitemShut
  {NoStop}%
\bibitem [{\citenamefont {Bian}\ \emph {et~al.}(2010)\citenamefont {Bian},
  \citenamefont {Chudak}, \citenamefont {Macready},\ and\ \citenamefont
  {Rose}}]{Bian2010}%
  \BibitemOpen
  \bibfield  {author} {\bibinfo {author} {\bibfnamefont {Z.}~\bibnamefont
  {Bian}}, \bibinfo {author} {\bibfnamefont {F.}~\bibnamefont {Chudak}},
  \bibinfo {author} {\bibfnamefont {W.}~\bibnamefont {Macready}}, \ and\
  \bibinfo {author} {\bibfnamefont {G.}~\bibnamefont {Rose}},\ }\bibfield
  {title} {\enquote {\bibinfo {title} {{The Ising model: teaching an old
  problem new tricks}},}\ }\href
  {https://www.dwavesys.com/sites/default/files/weightedmaxsat_v2.pdf}
  {\bibfield  {journal} {\bibinfo  {journal} {D-Wave Systems
  (www.dwavesys.com)}\ ,\ \bibinfo {pages} {1--32}} (\bibinfo {year}
  {2010})}\BibitemShut {NoStop}%
\bibitem [{\citenamefont {Johnson}\ \emph {et~al.}(2011)\citenamefont
  {Johnson}, \citenamefont {Amin}, \citenamefont {Gildert}, \citenamefont
  {Lanting}, \citenamefont {Hamze}, \citenamefont {Dickson}, \citenamefont
  {Harris}, \citenamefont {Berkley}, \citenamefont {Johansson}, \citenamefont
  {Bunyk}, \citenamefont {Chapple}, \citenamefont {Enderud}, \citenamefont
  {Hilton}, \citenamefont {Karimi}, \citenamefont {Ladizinsky}, \citenamefont
  {Ladizinsky}, \citenamefont {Oh}, \citenamefont {Perminov}, \citenamefont
  {Rich}, \citenamefont {Thom}, \citenamefont {Tolkacheva}, \citenamefont
  {Truncik}, \citenamefont {Uchaikin}, \citenamefont {Wang}, \citenamefont
  {Wilson},\ and\ \citenamefont {Rose}}]{Johnson2011}%
  \BibitemOpen
  \bibfield  {author} {\bibinfo {author} {\bibfnamefont {M.~W.}\ \bibnamefont
  {Johnson}}, \bibinfo {author} {\bibfnamefont {M.~H.~S.}\ \bibnamefont
  {Amin}}, \bibinfo {author} {\bibfnamefont {S.}~\bibnamefont {Gildert}},
  \bibinfo {author} {\bibfnamefont {T.}~\bibnamefont {Lanting}}, \bibinfo
  {author} {\bibfnamefont {F.}~\bibnamefont {Hamze}}, \bibinfo {author}
  {\bibfnamefont {N.}~\bibnamefont {Dickson}}, \bibinfo {author} {\bibfnamefont
  {R.}~\bibnamefont {Harris}}, \bibinfo {author} {\bibfnamefont {A.~J.}\
  \bibnamefont {Berkley}}, \bibinfo {author} {\bibfnamefont {J.}~\bibnamefont
  {Johansson}}, \bibinfo {author} {\bibfnamefont {P.}~\bibnamefont {Bunyk}},
  \bibinfo {author} {\bibfnamefont {E.~M.}\ \bibnamefont {Chapple}}, \bibinfo
  {author} {\bibfnamefont {C.}~\bibnamefont {Enderud}}, \bibinfo {author}
  {\bibfnamefont {J.~P.}\ \bibnamefont {Hilton}}, \bibinfo {author}
  {\bibfnamefont {K.}~\bibnamefont {Karimi}}, \bibinfo {author} {\bibfnamefont
  {E.}~\bibnamefont {Ladizinsky}}, \bibinfo {author} {\bibfnamefont
  {N.}~\bibnamefont {Ladizinsky}}, \bibinfo {author} {\bibfnamefont
  {T.}~\bibnamefont {Oh}}, \bibinfo {author} {\bibfnamefont {I.}~\bibnamefont
  {Perminov}}, \bibinfo {author} {\bibfnamefont {C.}~\bibnamefont {Rich}},
  \bibinfo {author} {\bibfnamefont {M.~C.}\ \bibnamefont {Thom}}, \bibinfo
  {author} {\bibfnamefont {E.}~\bibnamefont {Tolkacheva}}, \bibinfo {author}
  {\bibfnamefont {C.~J.~S.}\ \bibnamefont {Truncik}}, \bibinfo {author}
  {\bibfnamefont {S.}~\bibnamefont {Uchaikin}}, \bibinfo {author}
  {\bibfnamefont {J.}~\bibnamefont {Wang}}, \bibinfo {author} {\bibfnamefont
  {B.}~\bibnamefont {Wilson}}, \ and\ \bibinfo {author} {\bibfnamefont
  {G.}~\bibnamefont {Rose}},\ }\bibfield  {title} {\enquote {\bibinfo {title}
  {{Quantum annealing with manufactured spins}},}\ }\href
  {http://dx.doi.org/10.1038/nature10012
  http://www.nature.com/nature/journal/v473/n7346/abs/10.1038-nature10012-unlocked.html{\#}supplementary-information}
  {\bibfield  {journal} {\bibinfo  {journal} {Nature}\ }\textbf {\bibinfo
  {volume} {473}},\ \bibinfo {pages} {194} (\bibinfo {year}
  {2011})}\BibitemShut {NoStop}%
\bibitem [{\citenamefont {Messiah}(1966)}]{Messiah1966}%
  \BibitemOpen
  \bibfield  {author} {\bibinfo {author} {\bibfnamefont {A.}~\bibnamefont
  {Messiah}},\ }\href@noop {} {\emph {\bibinfo {title} {Quantum Mechanics}}},\
  Vol.~\bibinfo {volume} {II}\ (\bibinfo  {publisher} {John Wiley \& Sons},\
  \bibinfo {address} {Amsterdam, The Netherlands},\ \bibinfo {year}
  {1966})\BibitemShut {NoStop}%
\bibitem [{\citenamefont {Santos}\ and\ \citenamefont
  {Sarandy}(2015)}]{Santos2015}%
  \BibitemOpen
  \bibfield  {author} {\bibinfo {author} {\bibfnamefont {A.~C.}\ \bibnamefont
  {Santos}}\ and\ \bibinfo {author} {\bibfnamefont {M.~S.}\ \bibnamefont
  {Sarandy}},\ }\bibfield  {title} {\enquote {\bibinfo {title} {Superadiabatic
  controlled evolutions and universal quantum computation},}\ }\href
  {http://dx.doi.org/10.1038/srep15775} {\bibfield  {journal} {\bibinfo
  {journal} {Sci. Rep.}\ }\textbf {\bibinfo {volume} {5}},\ \bibinfo {pages}
  {15775} (\bibinfo {year} {2015})}\BibitemShut {NoStop}%
\bibitem [{\citenamefont {Santos}\ \emph {et~al.}()\citenamefont {Santos},
  \citenamefont {Silva},\ and\ \citenamefont {Sarandy}}]{Santos2016}%
  \BibitemOpen
  \bibfield  {author} {\bibinfo {author} {\bibfnamefont {A.~C.}\ \bibnamefont
  {Santos}}, \bibinfo {author} {\bibfnamefont {R.~D.}\ \bibnamefont {Silva}}, \
  and\ \bibinfo {author} {\bibfnamefont {M.~S.}\ \bibnamefont {Sarandy}},\
  }\bibfield  {title} {\enquote {\bibinfo {title} {Shortcut to adiabatic gate
  teleportation},}\ }\href {\doibase 10.1103/PhysRevA.93.012311} {\bibfield
  {journal} {\bibinfo  {journal} {Phys. Rev. A}\ }\textbf {\bibinfo {volume}
  {93}},\ \bibinfo {pages} {012311}}\BibitemShut {NoStop}%
\bibitem [{\citenamefont {{Coulamy}}\ \emph {et~al.}(2016)\citenamefont
  {{Coulamy}}, \citenamefont {{Santos}}, \citenamefont {{Hen}},\ and\
  \citenamefont {{Sarandy}}}]{Coulamy2016}%
  \BibitemOpen
  \bibfield  {author} {\bibinfo {author} {\bibfnamefont {I.~B.}\ \bibnamefont
  {{Coulamy}}}, \bibinfo {author} {\bibfnamefont {A.~C.}\ \bibnamefont
  {{Santos}}}, \bibinfo {author} {\bibfnamefont {I.}~\bibnamefont {{Hen}}}, \
  and\ \bibinfo {author} {\bibfnamefont {M.~S.}\ \bibnamefont {{Sarandy}}},\
  }\bibfield  {title} {\enquote {\bibinfo {title} {{Energetic cost of
  superadiabatic quantum computation}},}\ }\href@noop {} {\bibfield  {journal}
  {\bibinfo  {journal} {ArXiv e-prints}\ } (\bibinfo {year} {2016})},\ \Eprint
  {http://arxiv.org/abs/1603.07778} {arXiv:1603.07778 [quant-ph]} \BibitemShut
  {NoStop}%
\bibitem [{Note1()}]{Note1}%
  \BibitemOpen
  \bibinfo {note} {Interestingly a similar result was obtained in Ref.~\cite
  {Santos2015}. However, the discussion in Ref.~\cite {Santos2015} is based on
  a different version of the QSL \cite {Deffner2013}, which is governed by the
  off-diagonal matrix elements and not by the norm of $H_1(t)$ \protect \textup
  {\hbox {\mathsurround \z@ \protect \normalfont (\ignorespaces \ref
  {eq03}\unskip \@@italiccorr )}}. Therefore, the here discussed trade-off
  could not be analyzed previously.}\BibitemShut {Stop}%
\bibitem [{Note2()}]{Note2}%
  \BibitemOpen
  \bibinfo {note} {It is interesting to note that recently a similar relation
  was shown for work fluctuations induced by using TQD~\cite
  {Funo2016}.}\BibitemShut {Stop}%
\bibitem [{Note3()}]{Note3}%
  \BibitemOpen
  \bibinfo {note} {Note that in the unscaled case \protect \textup {\hbox
  {\mathsurround \z@ \protect \normalfont (\ignorespaces \ref {eq14}\unskip
  \@@italiccorr )}}, $v_\protect \mathrm {QSL}$ vanishes for the minimal, zero
  gap, cf. Eq.~\protect \textup {\hbox {\mathsurround \z@ \protect \normalfont
  (\ignorespaces \ref {eq07}\unskip \@@italiccorr )}}. After rescaling the
  minimal gap is finite, and hence also the speed remains finite. Therefore, we
  plot $\protect \qopname \relax o{log}(v_\protect \mathrm {QSL})$ to avoid
  misinterpretation of our findings due to rescaling.}\BibitemShut {Stop}%
\bibitem [{\citenamefont {Francuz}\ \emph {et~al.}(2016)\citenamefont
  {Francuz}, \citenamefont {Dziarmaga}, \citenamefont {Gardas},\ and\
  \citenamefont {Zurek}}]{Francuz2016}%
  \BibitemOpen
  \bibfield  {author} {\bibinfo {author} {\bibfnamefont {A.}~\bibnamefont
  {Francuz}}, \bibinfo {author} {\bibfnamefont {J.}~\bibnamefont {Dziarmaga}},
  \bibinfo {author} {\bibfnamefont {B.}~\bibnamefont {Gardas}}, \ and\ \bibinfo
  {author} {\bibfnamefont {W.~H.}\ \bibnamefont {Zurek}},\ }\bibfield  {title}
  {\enquote {\bibinfo {title} {Space and time renormalization in phase
  transition dynamics},}\ }\href {\doibase 10.1103/PhysRevB.93.075134}
  {\bibfield  {journal} {\bibinfo  {journal} {Phys. Rev. B}\ }\textbf {\bibinfo
  {volume} {93}},\ \bibinfo {pages} {075134} (\bibinfo {year}
  {2016})}\BibitemShut {NoStop}%
\bibitem [{\citenamefont {Bason}\ \emph {et~al.}(2012)\citenamefont {Bason},
  \citenamefont {Viteau}, \citenamefont {Malossi}, \citenamefont {Huillery},
  \citenamefont {Arimondo}, \citenamefont {Ciampini}, \citenamefont {Fazio},
  \citenamefont {Giovannetti}, \citenamefont {Mannella},\ and\ \citenamefont
  {Morsch}}]{Bason2012}%
  \BibitemOpen
  \bibfield  {author} {\bibinfo {author} {\bibfnamefont {M.~G.}\ \bibnamefont
  {Bason}}, \bibinfo {author} {\bibfnamefont {M.}~\bibnamefont {Viteau}},
  \bibinfo {author} {\bibfnamefont {N.}~\bibnamefont {Malossi}}, \bibinfo
  {author} {\bibfnamefont {P.}~\bibnamefont {Huillery}}, \bibinfo {author}
  {\bibfnamefont {E.}~\bibnamefont {Arimondo}}, \bibinfo {author}
  {\bibfnamefont {D.}~\bibnamefont {Ciampini}}, \bibinfo {author}
  {\bibfnamefont {R.}~\bibnamefont {Fazio}}, \bibinfo {author} {\bibfnamefont
  {V.}~\bibnamefont {Giovannetti}}, \bibinfo {author} {\bibfnamefont
  {R.}~\bibnamefont {Mannella}}, \ and\ \bibinfo {author} {\bibfnamefont
  {O.}~\bibnamefont {Morsch}},\ }\bibfield  {title} {\enquote {\bibinfo {title}
  {High-fidelity quantum driving},}\ }\href
  {http://dx.doi.org/10.1038/nphys2170} {\bibfield  {journal} {\bibinfo
  {journal} {Nat Phys}\ }\textbf {\bibinfo {volume} {8}},\ \bibinfo {pages}
  {147} (\bibinfo {year} {2012})}\BibitemShut {NoStop}%
\bibitem [{\citenamefont {Zhang}\ \emph {et~al.}(2013)\citenamefont {Zhang},
  \citenamefont {Shim}, \citenamefont {Niemeyer}, \citenamefont {Taniguchi},
  \citenamefont {Teraji}, \citenamefont {Abe}, \citenamefont {Onoda},
  \citenamefont {Yamamoto}, \citenamefont {Ohshima}, \citenamefont {Isoya},\
  and\ \citenamefont {Suter}}]{Zhang2013}%
  \BibitemOpen
  \bibfield  {author} {\bibinfo {author} {\bibfnamefont {J.}~\bibnamefont
  {Zhang}}, \bibinfo {author} {\bibfnamefont {J.~H.}\ \bibnamefont {Shim}},
  \bibinfo {author} {\bibfnamefont {I.}~\bibnamefont {Niemeyer}}, \bibinfo
  {author} {\bibfnamefont {T.}~\bibnamefont {Taniguchi}}, \bibinfo {author}
  {\bibfnamefont {T.}~\bibnamefont {Teraji}}, \bibinfo {author} {\bibfnamefont
  {H.}~\bibnamefont {Abe}}, \bibinfo {author} {\bibfnamefont {S.}~\bibnamefont
  {Onoda}}, \bibinfo {author} {\bibfnamefont {T.}~\bibnamefont {Yamamoto}},
  \bibinfo {author} {\bibfnamefont {T.}~\bibnamefont {Ohshima}}, \bibinfo
  {author} {\bibfnamefont {J.}~\bibnamefont {Isoya}}, \ and\ \bibinfo {author}
  {\bibfnamefont {D.}~\bibnamefont {Suter}},\ }\bibfield  {title} {\enquote
  {\bibinfo {title} {Experimental implementation of assisted quantum adiabatic
  passage in a single spin},}\ }\href {\doibase 10.1103/PhysRevLett.110.240501}
  {\bibfield  {journal} {\bibinfo  {journal} {Phys. Rev. Lett.}\ }\textbf
  {\bibinfo {volume} {110}},\ \bibinfo {pages} {240501} (\bibinfo {year}
  {2013})}\BibitemShut {NoStop}%
\bibitem [{\citenamefont {Campbell}(2016)}]{Campbell2016}%
  \BibitemOpen
  \bibfield  {author} {\bibinfo {author} {\bibfnamefont {S.}~\bibnamefont
  {Campbell}},\ }\bibfield  {title} {\enquote {\bibinfo {title} {Criticality
  revealed through quench dynamics in the {Lipkin-Meshkov-Glick} model},}\
  }\href {\doibase 10.1103/PhysRevB.94.184403} {\bibfield  {journal} {\bibinfo
  {journal} {Phys. Rev. B}\ }\textbf {\bibinfo {volume} {94}},\ \bibinfo
  {pages} {184403} (\bibinfo {year} {2016})}\BibitemShut {NoStop}%
\bibitem [{\citenamefont {{Vacanti}}\ \emph {et~al.}(2014)\citenamefont
  {{Vacanti}}, \citenamefont {{Fazio}}, \citenamefont {{Montangero}},
  \citenamefont {{Palma}}, \citenamefont {{Paternostro}},\ and\ \citenamefont
  {{Vedral}}}]{Vacanti2014}%
  \BibitemOpen
  \bibfield  {author} {\bibinfo {author} {\bibfnamefont {G.}~\bibnamefont
  {{Vacanti}}}, \bibinfo {author} {\bibfnamefont {R.}~\bibnamefont {{Fazio}}},
  \bibinfo {author} {\bibfnamefont {S.}~\bibnamefont {{Montangero}}}, \bibinfo
  {author} {\bibfnamefont {G.~M.}\ \bibnamefont {{Palma}}}, \bibinfo {author}
  {\bibfnamefont {M.}~\bibnamefont {{Paternostro}}}, \ and\ \bibinfo {author}
  {\bibfnamefont {V.}~\bibnamefont {{Vedral}}},\ }\bibfield  {title} {\enquote
  {\bibinfo {title} {{Transitionless quantum driving in open quantum
  systems}},}\ }\href {\doibase 10.1088/1367-2630/16/5/053017} {\bibfield
  {journal} {\bibinfo  {journal} {New J. Phys.}\ }\textbf {\bibinfo {volume}
  {16}},\ \bibinfo {eid} {053017} (\bibinfo {year} {2014})}\BibitemShut
  {NoStop}%
\bibitem [{\citenamefont {{Funo}}\ \emph {et~al.}(2016)\citenamefont {{Funo}},
  \citenamefont {{Zhang}}, \citenamefont {{Chatou}}, \citenamefont {{Kim}},
  \citenamefont {{Ueda}},\ and\ \citenamefont {{del Campo}}}]{Funo2016}%
  \BibitemOpen
  \bibfield  {author} {\bibinfo {author} {\bibfnamefont {K.}~\bibnamefont
  {{Funo}}}, \bibinfo {author} {\bibfnamefont {J.-N.}\ \bibnamefont {{Zhang}}},
  \bibinfo {author} {\bibfnamefont {C.}~\bibnamefont {{Chatou}}}, \bibinfo
  {author} {\bibfnamefont {K.}~\bibnamefont {{Kim}}}, \bibinfo {author}
  {\bibfnamefont {M.}~\bibnamefont {{Ueda}}}, \ and\ \bibinfo {author}
  {\bibfnamefont {A.}~\bibnamefont {{del Campo}}},\ }\bibfield  {title}
  {\enquote {\bibinfo {title} {{Universal Work Fluctuations during Shortcuts To
  Adiabaticity by Counterdiabatic Driving}},}\ }\href@noop {} {\bibfield
  {journal} {\bibinfo  {journal} {ArXiv e-prints}\ } (\bibinfo {year}
  {2016})},\ \Eprint {http://arxiv.org/abs/1609.08889} {arXiv:1609.08889
  [quant-ph]} \BibitemShut {NoStop}%
\end{thebibliography}%

\end{document}